\documentclass[preprint,aps,prd]{revtex4-1}
\usepackage{graphicx}
\usepackage{slashed}
\usepackage{verbatim} 
\newcommand{\be}{\begin{eqnarray}}
\newcommand{\ee}{\end{eqnarray}}
\usepackage[colorlinks]{hyperref}
\usepackage{cancel,bm} 
\usepackage{amsmath,amssymb}
\usepackage{mathtools}
\usepackage{float}
\usepackage{color}
\usepackage{leftidx}
\usepackage{booktabs}
\usepackage{latexsym}
\usepackage{revsymb}
\usepackage{multirow}
\usepackage{hypcap}
\usepackage{array}
\usepackage{subcaption}
\usepackage{slashed}
\usepackage[normalem]{ulem}
\usepackage{amsmath}
\newcommand{\stkout}[1]{\ifmmode\text{\sout{\ensuremath{#1}}}\else\sout{#1}\fi}

\hypersetup{
    colorlinks=red,
    citecolor=blue,
    filecolor=magenta,      
    urlcolor=red,
}
\renewcommand{\d}{\mathrm{d}}
\newcommand{\sT}{{\scriptscriptstyle T}}

\definecolor{Argya_color}{RGB}{200, 0, 200}
\definecolor{asmita_color}{rgb}{0.0, 0.1, 0.9}
\definecolor{rajesh_color}{rgb}{0.93, 0.53, 0.18}

\newcommand{\rajesh}[1]{{\color{rajesh_color} {#1}}}
\begin{document}

\title{The effect of TMD evolution on the Sivers asymmetry in back-to-back $J/\psi+\gamma$ and $J/\psi+\text{jet}$ production at the Electron-Ion-Collider}
\author{Arghya Jana}
 \email{23d1064@iitb.ac.in}
 \affiliation{Department of Physics, Indian Institute of Technology Bombay, Mumbai, Maharashtra 400076, India}
 
 \author{Asmita Mukherjee}
 \email{asmita@phy.iitb.ac.in}
 \affiliation{Department of Physics, Indian Institute of Technology Bombay, Mumbai, Maharashtra 400076, India}
 
 \author{Sangem Rajesh}
\email{sangem.rajesh@vit.ac.in}
 \affiliation{Department of Physics, School of Advanced Sciences, Vellore Institute of Technology, Vellore,
Tamil Nadu 632014, India}

\begin{abstract}
We present an estimate of the Sivers asymmetry in back-to-back $J/\psi$-photon and $J/\psi$-jet production in electron-proton collisions in the kinematics of the upcoming Electron-Ion Collider (EIC) in a transverse momentum dependent (TMD) factorization framework, and also incorporating the TMD evolution. We use the non-relativistic Quantum Chromodynamics (NRQCD) model to study the production mechanism of $J/\psi$. The gluon induced channel dominates, and these are promising probes of the less known gluon Sivers function. We incorporate the TMD evolution in the cross section and Sivers asymmetry in the Collins-Soper-Sterman (CSS) approach and show that the asymmetry is sizable even after the evolution. Although the cross section for $J/\psi$-jet production depends on the long-distance matrix element (LDME) set chosen, the asymmetry remains largely unaffected. The asymmetry is independent of the LDME at leading order for $J/\psi$-photon production. Thus, the Sivers asymmetry in both processes is a robust probe of the gluon Sivers function.   
\end{abstract}

\maketitle
\raggedbottom 

\section{Introduction}\label{sec:intro}
Transverse momentum dependent parton distribution functions (TMDs)~\cite{Mulders:1995dh,Boer:1997nt,Boer:1999uu,Anselmino:1999pw,Anselmino:1994tv,Barone:2001sp} play an important role in mapping the three-dimensional structure of hadrons. In the conventional collinear approach, parton distribution functions (PDFs) depend only on the longitudinal momentum fraction $x$ of the parton, while all information associated with its intrinsic transverse momentum $k_\perp$ is integrated out. As a result, the collinear framework provides only a one-dimensional description of the hadron structure. In contrast, TMDs are functions of both $x$ and $k_\perp$, thereby providing a three-dimensional picture of the internal structure of the hadron. These functions encode the spin-orbit correlation of partons inside the proton. Being non-perturbative objects, one can access TMDs experimentally in different processes like semi-inclusive deep inelastic scattering (SIDIS) and Drell-Yan (DY). An important feature of TMDs is their process dependence, which originates from the appearance of gauge links (Wilson lines) in their operator definitions~\cite{Mulders:2000sh}. These gauge links ensure color gauge invariance and resum contributions arising from initial and final state interactions. Unlike quark TMDs, which involve a single gauge link, gluon TMDs contain two gauge links, making them more strongly process dependent~\cite{Buffing:2013kca}.

Among eight independent gluon TMDs defined at leading twist, the gluon Sivers function (GSF)~\cite{Sivers:1989cc,Sivers:1990fh}, $f_{1T}^{\perp g}$, has attracted much attention in recent years. It describes how the unpolarized gluons are distributed inside a transversely polarized proton and arises from the interaction between the intrinsic transverse momentum of the gluon and the transverse spin of the proton. This correlation generates an asymmetric distribution of particles produced in the transverse plane, giving rise to the Sivers asymmetry~\cite{Sivers:1989cc,Sivers:1990fh}. Theoretical model calculations have suggested that the Sivers function contains information about the orbital motion of partons~\cite{Burkardt:2003uw,Burkardt:2003je}. Its first transverse moment has a direct relation with the twist-three Qiu--Sterman function~\cite{Qiu:1991pp,Qiu:1998ia}. It is a time-reversal odd distribution arising due to the presence of process dependent gauge links and is predicted to have opposite signs in SIDIS and DY processes~\cite{Brodsky:2002rv,Collins:2002kn,Gamberg:2013kla,Kang:2011hk}. Recent measurements from RHIC~\cite{STAR:2015vmv} and COMPASS~\cite{COMPASS:2017jbv} support this predicted sign change, although more data are still required for a definite confirmation~\cite{Anselmino:2016uie}. In general, the GSF can be decomposed into two independent components, called the $f$-type (Weizs\"acker-Williams) and $d$-type (dipole) GSF~\cite{Collins:2002kn,Brodsky:2002cx,Belitsky:2002sm,Ji:2002aa,Boer:2003cm}, the operator structure and the gauge links involved are different in these two cases. The first experimental indications of a nonzero quark Sivers function were obtained from the measurements of the HERMES~\cite{HERMES:2004mhh,HERMES:2009lmz} and COMPASS~\cite{COMPASS:2012dmt,JeffersonLabHallA:2011ayy} collaborations. The quark Sivers function has been extracted in Ref.~\cite{Anselmino:2016uie}. Although an extraction of gluon Sivers function is still not available, it has been studied in Refs.~\cite{DAlesio:2015fwo,DAlesio:2018rnv}, by fitting RHIC data within the generalized parton model (GPM). Ref.~\cite{Echevarria:2026vca} has recently studied the Sivers asymmetry for dijet production in SIDIS within the TMD factorization framework incorporating TMD evolution effects and presented asymmetries in the kinematics of the future Electron-Ion Collider(EIC). The TMD evolution, which governs the scale dependence of TMDs, differs from the evolution of collinear PDFs and is described by the Collins-Soper-Sterman (CSS) formalism~\cite{Collins:1984kg,Qiu:2000ga,Landry:2002ix}. In~\cite{Echevarria:2014xaa}, the Sivers function was parametrized by taking TMD evolution effects into account. Nevertheless, the GSF is much less known than the quark Sivers function. Gluon TMDs satisfy positivity bounds~\cite{Mulders:2000sh}, and moreover they are constrained by the Burkardt sum rule~\cite{Burkardt:2004ur}. Analyses of SIDIS data at low energy scale~\cite{Anselmino:2008sga} indicate that the dominant contributions to the sum rule arise from up and down quarks, although a remaining contribution of about $30\%$ may still arise from the GSF. As a result, gluon TMDs remain largely uncertain and will be a key focus of future studies at the planned EIC~\cite{Accardi:2012qut}.

Transverse momentum dependent (TMD) evolution describes how TMDs change with the energy scale of the process. In general, TMDs exhibit dependence on both the renormalization scale and the rapidity scale. The scale dependence is described by the renormalization group and Collins--Soper equations~\cite{Aybat:2011ge,Aybat:2011zv,Collins:2011zzd}. The evolution is naturally performed in impact parameter ($b_T$) space, where the TMDs are expressed using a Fourier transform. In this space, the evolution is controlled by the Sudakov factor. The perturbative Sudakov factor works well in the small-$b_T$ region. At large $b_T$, the perturbation theory is not applicable. Therefore, the $b_\ast$ prescription~\cite{Aybat:2011ge,Aybat:2011zv,Collins:2011zzd} is used so that the calculation remains in the perturbative region. The non-perturbative effects at large $b_T$ are included through a non-perturbative Sudakov factor, which is usually extracted from the data. Therefore, the TMD evolution provides a consistent framework to connect different energy scales. It properly accounts for both perturbative and non-perturbative effects. As a result, TMD evolution is important for determining the realistic transverse momentum dependence of TMDs and predicting physical observables reliably. 

The $J/\psi$ meson, the lightest bound state of a charm quark and an anti-charm quark, provides a useful probe of gluon dynamics in high-energy collisions. In the present work, we employ the non-relativistic Quantum Chromodynamics (NRQCD) framework~\cite{Bodwin:1994jh}, which is widely used for describing heavy quarkonium production and decay. It provides a rigorous and systematic description using an effective field theory approach. Within this framework, one can factorize the cross section of quarkonium (e.g., $J/\psi$) production into a hard part and a soft part. The hard part corresponds to the formation of a heavy quark-antiquark ($c\bar{c}$) pair in a hard scattering process and can be calculated using perturbative QCD~\cite{Boer:2012bt}. The formation of the physical $J/\psi$ from the $c\bar{c}$ pair through soft gluon radiation is described by the soft part. This non-perturbative transition is characterized by long-distance matrix elements (LDME)~\cite{Bodwin:1994jh}. These LDMEs are expected to be process independent and are typically extracted by fitting experimental data. In NRQCD, the heavy quark-antiquark pair may be formed in different intermediate states characterized by the spectroscopic notation $^{2S+1}L_J^{(1,8)}$, where $S$, $L$, and $J$ denote the spin, orbital, and total angular momentum of the pair, respectively, and the superscripts $(1)$ and $(8)$ represent the color singlet (CS) and color octet (CO) configurations, respectively. In this model, the theory is systematically expanded in powers of the strong coupling constant $\alpha_s$ and the relative velocity $v$ of the heavy quark in the quarkonium rest frame. In the non-relativistic limit $v\ll 1$, the LDMEs scale with definite powers of $v$. For heavy quarkonium systems such as charmonium and bottomonium, the typical values are $v^2 \approx 0.3$ and $v^2 \approx 0.1$, respectively.

The GSF has been probed in various processes. In electron-proton ($ep$) collision, the Sivers asymmetry is generated by the virtual photon--gluon scattering. Ref.~\cite{Mukherjee:2016qxa} investigated the Sivers asymmetry for the electroproduction of $J/\psi$ via the leading-order subprocess $\gamma^\ast\, g \rightarrow J/\psi$, including the TMD evolution effect. The GSF has also been studied through $J/\psi$ photoproduction~\cite{Godbole:2012bx,Godbole:2013bca,Rajesh:2018qks}. Azimuthal asymmetries for the electroproduction of $J/\psi$--jet were studied in back-to-back arrangement in Ref.~\cite{DAlesio:2019qpk}, where the upper bounds of the asymmetries were estimated. In Ref.~\cite{Chakrabarti:2022rjr}, the Sivers asymmetry has been examined in $J/\psi$--photon electroproduction channel in back-to-back kinematics using the GPM, and estimates of the azimuthal asymmetries are provided by imposing the positivity constraint on TMDs. The Sivers asymmetry in $J/\psi$--jet photoproduction channel with back-to-back kinematics was further investigated in Ref.~\cite{Kishore:2019fzb} by incorporating TMD evolution. The GSF has also been explored in proton-proton ($pp$) collision, where the asymmetry arises from the fusion of two gluons~\cite{DAlesio:2017rzj}.

In this paper, we investigate the Sivers asymmetry in electroproduction of back-to-back $J/\psi$--photon and $J/\psi$--jet within the TMD evolution approach. The processes, considered in this article, are handy tools to probe the unknown GSF. Theoretical predictions are estimated to be relevant for the kinematics of the future EIC. We organize the paper as follows. We begin with an introduction in Sec.~\ref{sec:intro}. Sec.~\ref{sec:formalism} contains the formalism, while Sec.~\ref{sec:Sivers} discusses the Sivers asymmetry. The TMD evolution is explained in Sec.~\ref{sec:evolution}, and the numerical results are shown in Sec.~\ref{sec:results}. Finally, we summarize our conclusions in Sec.~\ref{sec:conclusion}.

\section{Formalism}
\label{sec:formalism}
We consider the production of a $J/\psi$ meson together with a photon or a jet in the semi-inclusive electron–proton scattering process
\begin{equation}
e(l) + p^\uparrow(P) \to e(l') + J/\psi(P_\psi) + \gamma(p_\gamma)/\mathrm{jet}(p_j) + X\,.
\end{equation}
The particle four-momenta are given in parentheses, and the arrow indicates the proton's transverse polarization. \\
For the final-state photon, at leading order (LO) we consider only the gluon channel $\gamma^{\ast} + g \to J/\psi + \gamma$, while the quark channel is neglected since it contributes at higher order. The gluon initiated process happens through the CO mechanism. At LO, an additional photon initiated process $\gamma^{\ast} + \gamma \to J/\psi + \gamma$, where the photon is emitted elastically or inelastically from the proton, contributes via CS mechanism. However, this contribution is neglected in the present work since the gluon density inside the proton is significantly larger than the photon density as discussed in Ref.~\cite{Kniehl:2006qq}. Although, the CO LDME is suppressed by $\mathcal{O}(v^4)$ relative to the CS one, this suppression is much smaller than the enhancement from the large gluon density in the proton, and therefore the CO channel can still contribute.

On the other hand, for the final-state jet we consider both the gluon channel $\gamma^{\ast} + g \to J/\psi + g$ and the quark\,(anti-quark) channel $\gamma^{\ast} + q\,(\bar{q}) \to J/\psi + q\,(\bar{q})$. However, the quark\,(anti-quark) channel contribution is found to be much smaller than the gluon channel for the kinematics considered in this work. \\
We work in the center-of-mass (cm) frame of the proton and virtual photon, where both are considered to move in the $z$-direction. Their four-momenta, $P^\mu$ and $q^\mu$, can be written as follows:
\begin{align}
P^\mu &= n_-^\mu + \frac{M_p^2}{2} n_+^\mu \approx n_-^\mu, \\
q^\mu &= -x_B n_-^\mu + \frac{Q^2}{2 x_B} n_+^\mu \approx -x_B P^\mu + (P \cdot q) n_+^\mu,\label{q_momenta}
\end{align}
where, $n_+$ and $n_-$ are light-like vectors ($n_+^2 = 0, n_-^2 = 0$, $n_+ \cdot n_- = 1$). $Q^2=-q^2$ is the virtuality of the photon. $x_B = \frac{Q^2}{2 P \cdot q}$ is the Bjorken variable. The proton mass is represented by $M_p$.\\
The electron-proton cm energy is
\begin{equation}
s = (P + l)^2 \approx 2 P \cdot l = \frac{2 P \cdot q}{y},
\end{equation}
where $y = \frac{P \cdot q}{P \cdot l}$ denotes the fraction of the electron's energy carried by the photon. The invariant mass squared for the system of a virtual photon and proton is given by
\begin{equation}
W^2 = (q + P)^2 \approx Q^2 \frac{1-x_B}{x_B} \approx y s - Q^2.
\end{equation}
The four-momenta of incoming and scattered electrons are written as
\begin{align}
l^\mu &= \frac{1-y}{y} x_B\,n_-^\mu + \frac{1}{y}\frac{Q^2}{2 x_B} n_+^\mu + \frac{\sqrt{1-y}}{y} \, Q\, \hat{l}_\perp^\mu,\\
l^{\prime\mu} &= l^\mu - q^\mu \nonumber\\
&=\frac{1}{y} x_B\,n_-^\mu + \frac{1-y}{y}\frac{Q^2}{2x_B} n_+^\mu + \frac{\sqrt{1-y}}{y} \, Q\, \hat{l}_\perp^\mu,
\end{align}
where $\hat{l}_\perp^\mu$ is the unit vector in the transverse plane. We choose the lepton scattering plane, defined by these momenta, as the reference plane in which the azimuthal angles satisfy $\phi_l = \phi_{l'} = 0$. All angles are measured with respect to this plane.

Within the TMD factorization framework, the differential cross section can be expressed as a convolution of the leptonic tensor, the parton correlator, and the squared partonic hard-scattering amplitude~\cite{Pisano:2013cya}
\begin{align}
\d\sigma &=\sum_{a,a^\prime} \frac{1}{2s} \frac{\d^3 \bm l'}{(2\pi)^3 2 E_{l'}} \frac{\d^3 \bm P_\psi}{(2\pi)^3 2 E_\psi} \frac{\d^3 \bm p_{a^\prime}}{(2\pi)^3 2 E_{a^\prime}}  \int \d x \, \d^2 \bm p_{\sT} \, (2\pi)^4 \delta^4(q + p_a - P_\psi - p_{a'}) \nonumber\\
&\quad \times\frac{1}{Q^4} 
L(l,q)\otimes \Phi_a(x, \bm p_{\sT}) \otimes \Big|H_{\gamma^{\ast}a \to J/\psi \,a'}(q,p_a,P_\psi,p_{a'})\Big|^2\,,\label{dsigma}
\end{align}
where $a$ denotes the initial-state parton and $a'$ denotes the final-state particle. Here, $a=g$ and $a'=\gamma$ for $J/\psi + \gamma$ production, while for $J/\psi+\mathrm{jet}$ production, $a=a'=g, q, \bar{q}$. The convolution $\otimes$ represents the appropriate traces over Dirac indices.

The leptonic tensor, describing the electron-virtual photon interaction, is given by
\begin{equation}
L^{\mu\nu} = e^2 \left[ - g^{\mu\nu} Q^2 + 2 \left( l^\mu l'^{\nu} + l^{\nu} l'^\mu\right) \right],
\end{equation}
where $e$ is the charge of an electron.\\
Using the expressions of $l$ and $l'$, the above equation can be rewritten as
\begin{align}
L^{\mu\nu} &= \frac{e^2 Q^2}{y^2} \Bigg[
- (2-2y+y^2) g_T^{\mu\nu} 
+ 4(1-y) \epsilon_L^\mu \epsilon_L^{\nu}
+ 4(1-y) \Big(\hat{l}_\perp^\mu \hat{l}_\perp^{\nu} + \frac{1}{2} g_T^{\mu\nu}\Big)\nonumber\\
&\quad  
+ 2 \sqrt{1-y}(2-y) \big(\epsilon_L^\mu \hat{l}_\perp^{\nu} + \epsilon_L^{\nu} \hat{l}_\perp^\mu \big)
\Bigg],
\end{align}
where the transverse projection tensor is given by
\begin{align}
g_T^{\mu\nu}&= g^{\mu\nu} - n_+^\mu n_-^\nu - n_+^\nu n_-^\mu
=g^{\mu\nu} - \frac{1}{P \cdot q}(P^\mu q^\nu + P^\nu q^\mu) - \frac{Q^2}{(P\cdot q)^2} P^\mu P^\nu ,
\end{align}
and the virtual photon's longitudinal polarization vector is defined as
\begin{equation}
\epsilon_L^\mu(q) = \frac{1}{Q} \Big(q^\mu + \frac{Q^2}{P \cdot q} P^\mu\Big),
\end{equation}
satisfying $\epsilon_L^2 = 1$ and $\epsilon_L \cdot q = 0$.

The parton correlators, i.e., the gluon and quark correlators, which are non-perturbative objects, encode the distribution of gluons and quarks within the proton, respectively. \\ 
For an unpolarized proton, the parametrization of the gluon correlator can be written in terms of the unpolarized gluon TMD $f_1^g$ as \cite{Mulders:2000sh,Meissner:2007rx}
\begin{equation}
\Phi_{g}^{U\,\mu\nu}(x,\bm p_{\sT}) =- \frac{1}{2x} 
  g_T^{\mu\nu} f_1^g(x,\bm p_{\sT}^2).\label{unp_cor_glu}
\end{equation}
For a transversely polarized proton characterized by transverse spin $\bm S_\sT$, its parametrization is given in terms of the gluon Sivers function $f_{1T}^{\perp g}$ as \cite{Mulders:2000sh,Meissner:2007rx}
\begin{align}
\Phi_{g}^{T\,\mu\nu}(x,\bm p_{\sT}) &= -\frac{1}{2x} \,g_T^{\mu\nu} \, \frac{\epsilon_T^{\alpha\beta} p_{T \alpha} S_{T \beta}}{M_p} \, f_{1T}^{\perp g}(x,\bm p_{\sT}^2),\label{trans_cor_glu}
\end{align}
where $\epsilon_T^{\mu\nu} = \epsilon^{\mu\nu\alpha\beta} n_{-\alpha} n_{+\beta}$ is the antisymmetric tensor with $\epsilon_T^{12}=+1$.\\
The quark correlator for an unpolarized proton can be parametrized in terms of  the unpolarized quark TMD $f_1^q$ as \cite{Boer:1997nt,Mulders:1995dh}
\begin{equation}
\Phi^U_{q}(x,\bm p_{\sT})=\frac{1}{2x} f_1^q(x,\bm p_{\sT}^2)\,\slashed{n}_+.\label{unp_cor_quark}
\end{equation}
For a transversely polarized proton, its parametrization is expressed using the quark Sivers function $f_{1T}^{\perp q}$ as \cite{Boer:1997nt,Mulders:1995dh}
\begin{equation}
\Phi^T_{q}(x,\bm p_{\sT})
=\frac{1}{2x}\,\frac{\epsilon^{\mu\nu\alpha\beta}\gamma_\mu
n_{+\nu}
p_{T\alpha} S_{T\beta}}{M_p} \,f_{1T}^{\perp q}(x,\bm p_{\sT}^2).\label{trans_cor_quark}
\end{equation}
For the present work, in Eqs.~\eqref{unp_cor_glu} and \eqref{unp_cor_quark}, we include only the contributions from the unpolarized gluon TMD $f_{1}^{g}$ and the unpolarized quark TMD $f_{1}^{q}$, respectively. In Eqs.~\eqref{trans_cor_glu} and \eqref{trans_cor_quark}, we include only the contributions from $f_{1T}^{\perp g}$ and $f_{1T}^{\perp q}$, corresponding to the gluon and quark Sivers functions, respectively. The TMDs $f_1^g$ and $f_1^q$ describe the unpolarized gluon and quark densities associated with an unpolarized proton, respectively, while the Sivers functions $f_{1T}^{\perp g}$ and $f_{1T}^{\perp q}$ describe the distribution of unpolarized gluons and quarks inside a transversely polarized proton, respectively.

The hard part $H$ represents the scattering amplitude for the partonic process $\gamma^{\ast}(q) + a(p_a) \to J/\psi(P_\psi) + a'(p_{a'})$. \\
The four-momenta of the initial parton, $J/\psi$, and the outgoing photon\,(jet) can be written as
\begin{align}
p_a^\mu &= x P^\mu + (p_a \cdot P -M_p^2\,x)n_+^\mu + p_{\sT}^\mu \approx x P^\mu + p_{\sT}^\mu, \nonumber\\
P_\psi^\mu &= z (P \cdot q) n_+^\mu + \frac{M_\psi^2 + \bm P_{\psi \perp}^2}{2 z P \cdot q} P^\mu + P_{\psi \perp}^\mu, \nonumber\\
p_{a'}^{\mu} &= (1-z)(P \cdot q) n_+^\mu + \frac{\bm{p}_{a'\perp}^{2}}{2(1-z) P \cdot q} P^\mu + p_{a' \perp}^{\mu}.\label{4 momenta}
\end{align}
Here, $x=p_a \cdot n_+$ is the longitudinal momentum fraction carried by the initial parton, whereas $p_\sT$ is its intrinsic transverse momentum. $z = \frac{P \cdot P_\psi}{P \cdot q}$ is an inelastic variable, representing the fraction of the virtual photon's energy carried by $J/\psi$. The transverse momenta of the $J/\psi$ and the outgoing photon\,(jet) are denoted by $\bm{P}_{\psi\perp}$ and $\bm{p}_{a'\perp}$, respectively. $M_\psi$ is the mass of the $J/\psi$. \\ 
Now, we define the Mandelstam variables as
\begin{align}
\hat{s} &= (q+p_a)^2 = Q^2 \, \frac{x - x_B}{x_B}, \nonumber\\
\hat{t} &= (q - P_\psi)^2 = \frac{1}{z}(z-1)(z Q^2 + M_\psi^2) - \frac{1}{z} \bm P_{\psi \perp}^2, \nonumber\\
\hat{u} &= (p_a - P_\psi)^2 = M_\psi^2 - \frac{x z}{x_B} Q^2.\label{mandelstam}
\end{align}
Using the decomposition of $q$ from Eq.~\eqref{q_momenta} and $p_a,\,P_\psi,\,p_{a'}$ from Eq.~\eqref{4 momenta}, the delta function in Eq.~\eqref{dsigma}, which ensures the four-momentum conservation, can be simplified as
\begin{align}
\delta^4(q+p_a-P_\psi-p_{a'}) &= \frac{2}{ys} \, \delta\!\left(1-z - \bar{z}\right)
\delta\!\left(x - \frac{\bar{z}(M_\psi^2 + \bm P_{\psi\perp}^2) + z \bm p_{a'\perp}^2 + z\bar{z} Q^2}{z(1-z)ys}\right) \nonumber\\
&\quad \times \delta^2(\bm p_{\sT} - \bm P_{\psi\perp} - \bm p_{a'\perp}),\label{del}
\end{align}
where $\bar{z} = 1-z$.\\
The phase space for the final-state particles can be written as
\begin{align}
\frac{\d^3 \bm l'}{(2\pi)^3 2E_{l'}} &= \frac{1}{16\pi^2} \, \d Q^2 \d y, \nonumber\\
\frac{\d^3 \bm P_\psi}{(2\pi)^3 2E_\psi} &= \frac{\d^2 \bm P_{\psi\perp} \, \d z}{(2\pi)^3 \, 2z}, \nonumber\\
\frac{\d^3 \bm p_{a'}}{(2\pi)^3 2E_{a'}} &= \frac{\d^2 \bm p_{a'\perp} \, \d\bar{z}}{(2\pi)^3 \, 2\bar{z}}.\label{phase}
\end{align}
Substituting Eqs.~\eqref{del} and \eqref{phase} into Eq.~\eqref{dsigma} and then integrating over $\bar{z}$, $x$, and $\bm p_\sT$, we get
\begin{align}
\d\sigma &= \frac{\d z \, \d y \, \d Q^2 \, \d^2 \bm q_{\sT} \, \d^2 \bm K_\perp}{(2\pi)^4} 
\frac{1}{16 y s^2 z(1-z) Q^4} \nonumber\\
&\quad \times \sum_{a,a^\prime}
L(l,q)\otimes \Phi_a(x, \bm q_{\sT}) \otimes \Big|H_{\gamma^{\ast}a \to J/\psi \,a'}(q,p_a,P_\psi,p_{a'})\Big|^2,\label{no_cont}
\end{align}
where $\bm q_{\sT}$ and $\bm{K}_{\perp}$ are defined as the total and half the relative transverse momentum of the outgoing particles, respectively.
\begin{equation}
\bm{q}_{\sT} = \bm{P}_{\psi\perp} + \bm{p}_{a'\perp}\,, 
\qquad \bm{K}_{\perp} = \frac{1}{2}(\bm{P}_{\psi\perp} - \bm{p}_{a'\perp}).
\end{equation}

\section{Sivers asymmetry in $J/\psi$–photon and $J/\psi$–jet production}
\label{sec:Sivers}
\begin{figure}[h]
\begin{center} 
\includegraphics[height=6cm,width=17cm]{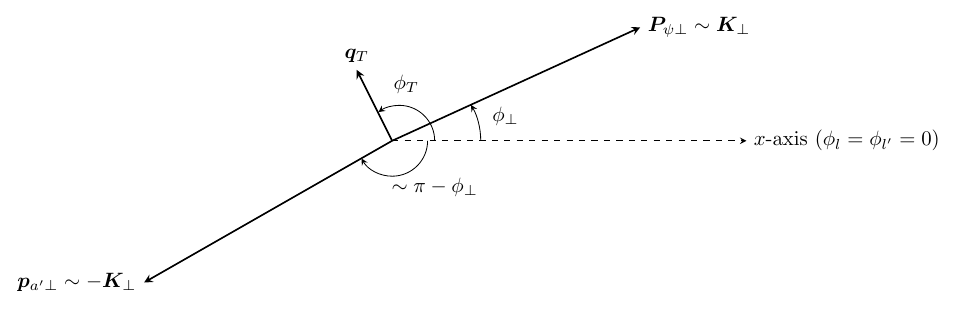}
\end{center}
\caption{\label{fig:1}Illustration of azimuthal angles in $J/\psi+\gamma\,(\mathrm{jet})$ production in semi-inclusive DIS.}
\end{figure}

As discussed above, we are interested in the kinematical region where $|\bm{q}_\sT| \ll |\bm{K}_\perp|$, as this kinematics allows the use of TMD factorization for the processes under consideration. 
The final-state particles, $J/\psi$ and the photon(jet), are produced nearly in opposite directions in the plane perpendicular to the proton and virtual photon directions, as shown in Fig.~\ref{fig:1}. \\
Therefore, we can approximately write
\[
\bm{P}_{\psi \perp} \simeq -\bm{p}_{a' \perp} \simeq \bm{K}_\perp.
\]
After contracting the leptonic tensor, the parton correlator, and the hard part, Eq.~\eqref{no_cont} reduces to
\begin{equation}
  \frac{\d\sigma}{\d z\,\d y\,\d Q^2\,\d ^2 \bm q_\sT\, \d ^2 \bm K_\perp}
  \equiv \d\sigma(\phi_S, \phi_T, \phi_\perp)
  = \d\sigma^U(\phi_T, \phi_\perp)
  + \d\sigma^T(\phi_S, \phi_T, \phi_\perp),
  \label{eq:dsigma-decomp}
\end{equation}
where $\d\sigma^U$ and $\d\sigma^T$ denote the cross sections corresponding to an unpolarized and a transversely polarized proton, respectively.\\
For $J/\psi+\gamma$ production, at leading order only the gluon channel contributes. The unpolarized and transversely polarized cross sections are given by
\begin{align}
  \d\sigma^U &=
  N \Big( A_0^g + A_1^g \cos \phi_\perp + A_2^g \cos 2\phi_\perp \Big)
  f_1^g(x,\bm q_{\sT}^2),
  \label{eq:dsigma-U-gamma}\\
  \d\sigma^T &= 
  N\,|\bm{S}_\sT|\,
  \sin(\phi_S - \phi_T)\,
  \Big( A_0^g + A_1^g \cos \phi_\perp + A_2^g \cos 2\phi_\perp \Big)\,
  \frac{q_\sT}{M_p}\,
  f_{1T}^{\perp g}(x,\bm q_{\sT}^2) .
  \label{eq:dsigma-T-gamma}
\end{align}
For $J/\psi+\mathrm{jet}$ production, both gluon and quark\,(anti-quark) channels contribute. The corresponding cross sections can be written as
\begin{align}
  \d\sigma^U &=
  N \sum_{a=g,q,\bar{q}}\Big( \mathcal{A}_0^a + \mathcal{A}_1^a \cos \phi_\perp + \mathcal{A}_2^a \cos 2\phi_\perp \Big)
  f_1^a(x,\bm q_{\sT}^2),
  \label{eq:dsigma-U-jet}\\
  \d\sigma^T &= 
  N\,|\bm{S}_\sT|\,
  \sin(\phi_S - \phi_T)\,
  \sum_{a=g,q,\bar{q}}
  \Big( \mathcal{A}_0^a + \mathcal{A}_1^a \cos \phi_\perp + \mathcal{A}_2^a \cos 2\phi_\perp \Big)\,
  \frac{q_\sT}{M_p}\,
  f_{1T}^{\perp a}(x,\bm q_{\sT}^2) .
  \label{eq:dsigma-T-jet}
\end{align}
where $N$ is given by
\begin{equation*}
  N = \frac{1}{(2\pi)^4 \, 16\, x\,y s^2\, z(1-z)\, Q^4}\,.
  \label{eq:N-factor}
\end{equation*}
The azimuthal angles corresponding to $\bm{q}_\sT$, $\bm{S}_\sT$, and 
$\bm{K}_\perp$ are denoted by $\phi_T$, $\phi_S$, and $\phi_\perp$, respectively, as shown in Fig.~\ref{fig:1}.\\ 
Here, $A_i^g$ and $\mathcal{A}_i^a$ ($i=0,1,2$) represent the amplitude modulations for the $J/\psi+\gamma$ and $J/\psi+\mathrm{jet}$ processes, respectively, where $a=g,q,\bar{q}$. For our calculation, only $A_0$($\mathcal{A}_0$) is required, whose expression agrees with the results of Refs.~\cite{DAlesio:2019qpk,Chakrabarti:2022rjr}.

The azimuthal modulations in Eqs.~\eqref{eq:dsigma-U-gamma}--\eqref{eq:dsigma-T-jet} can be used to isolate specific gluon TMDs through weighted asymmetries.  
The weighted asymmetry is given by~\cite{DAlesio:2019qpk}
\begin{equation}
  A^{W(\phi_S , \phi_T)} \equiv 
  2 \, \frac{\displaystyle \int \d\phi_S \, \d\phi_T \, \d\phi_\perp \,
  W(\phi_S , \phi_T)\, \d\sigma(\phi_S , \phi_T , \phi_\perp)}
  {\displaystyle \int \d\phi_S \, \d\phi_T \, \d\phi_\perp \,
  \d\sigma(\phi_S , \phi_T , \phi_\perp)} \, ,
  \label{eq:Sivers}
\end{equation}
where $W(\phi_S , \phi_T)$ denotes the weight function.  

In this study, we focus on the Sivers asymmetry, $A^{\sin(\phi_S-\phi_T)}$, which provides access to the GSF.
For the $J/\psi+\gamma$ process, substituting Eqs.~\eqref{eq:dsigma-U-gamma} and \eqref{eq:dsigma-T-gamma} into Eq.~\eqref{eq:Sivers}, and using the weight function $W(\phi_S , \phi_T)=\sin(\phi_S-\phi_T)$, we obtain the asymmetry as a function of $z$, $y$, $Q^2$, $q_\sT$ and $K_\perp$.
\begin{equation}
A^{\sin(\phi_S-\phi_T)} =
\frac{q_\sT}{M_p}  \,
\frac{f^{\perp g}_{1T}(x,\bm q_{\sT}^2)}{f^g_1(x,\bm q_{\sT}^2)}\, .
\label{ratio_gamma}
\end{equation}
For the $J/\psi+\mathrm{jet}$ production, the asymmetry becomes
\begin{equation}
A^{\sin(\phi_S-\phi_T)} =
\frac{q_\sT}{M_p}  \,
\frac{\displaystyle \sum_{a=g,q,\bar{q}} \mathcal{A}_0^a\,f^{\perp a}_{1T}(x,\bm q_{\sT}^2)}{\displaystyle \sum_{a=g,q,\bar{q}} \mathcal{A}_0^a\,f^{a}_{1}(x,\bm q_{\sT}^2)} \,.
\label{ratio_jet}
\end{equation}
For the $J/\psi+\gamma$ process, the asymmetry is independent of amplitude modulation factor $A_0^g$ due to its cancellation in the numerator and denominator. However, for the $J/\psi+\mathrm{jet}$ process, such a cancellation does not occur as multiple quark and gluon initiated processes contribute, and the amplitude modulation is different for each such process.
\section{Evolution of TMDs}
\label{sec:evolution}
In this section, we study the TMD evolution in the CSS formalism~\cite{Collins:1984kg,Qiu:2000ga,Landry:2002ix}. Usually, the evolution is studied in the impact parameter space. One defines TMDs in impact parameter ($b_T$) space as
\begin{align}
\tilde{f}_{1}^{g}(x,\boldsymbol{b}_{\sT}^2;\mu) 
&=  \int \d^2\boldsymbol{q}_{\sT} \, e^{-i \boldsymbol{b}_{\sT} \cdot \boldsymbol{q}_{\sT}} 
\, {f}_{1}^{ g}(x,\boldsymbol{q}_{\sT}^2;\mu) \,
\end{align}
and the inverse Fourier transform is given by
\begin{align}
f_{1}^{g}(x,\boldsymbol{q}_{\sT}^2;\mu) 
&= \frac{1}{(2\pi)^2} \int \d^2\boldsymbol{b}_{\sT} \, e^{i \boldsymbol{b}_{\sT} \cdot \boldsymbol{q}_{\sT}} 
\, \tilde{f}_{1}^{ g}(x,\boldsymbol{b}_{\sT}^2;\mu) \nonumber \\
&= \frac{1}{2\pi} \int_0^{\infty} \d b_{\sT} \, b_{\sT} \, J_0(b_{\sT} q_{\sT}) 
\, \tilde{f}_{1}^{ g}(x, \bm b_{\sT}^2; \mu)\,,\label{Fourier}
\end{align}
where $\bm{b}_{\sT}$ is the impact parameter in transverse coordinate space, Fourier conjugate to the transverse momentum $\bm{q}_{\sT}$, with $b_{\sT} \equiv |\bm{b}_{\sT}|$, and $J_0$ denotes the zeroth-order Bessel function of the first kind.

In general, TMDs are characterized by two scales: the renormalization scale $\mu$ and the rapidity scale $\zeta$. The former is introduced to remove the ultraviolet (UV) divergences arising from high-energy (short-distance) contributions, while the latter is required to regulate rapidity (light-cone) divergences associated with Wilson lines along light-like directions.

The evolution of TMDs with respect to $\mu$ and $\zeta$ is governed by the renormalization group (RG) and Collins--Soper (CS) equations~\cite{Aybat:2011ge,Aybat:2011zv,Collins:2011zzd}. The solutions of these equations determine the scale dependence of TMDs as they evolve from the initial scale 
\[
\mu_b = \frac{b_0}{b_\sT}, \qquad b_0 = 2 e^{-\gamma_E},
\]
where $\gamma_E = 0.577$ is the Euler-Mascheroni constant,
to the final hard scale $\mu = \sqrt{M_\psi^2 + Q^2}$.
The unpolarized gluon TMD $\tilde{f}_1^g$, in the perturbative regime, $b_\sT \ll 1/\Lambda_{\mathrm{QCD}}$, can be written as\,\cite{Boer:2023zit,Aybat:2011ge,Echevarria:2014xaa}
\begin{align}
    \tilde{f}_{1}^{g}\left(x, \boldsymbol{b}_{\sT}^2 ; \mu_b,\mu\right)
    &= \sum_{a = g, q, \bar{q}} 
    \left(C_{g / a} \otimes f_{1}^{a}\right)\left(x, \mu_{b}\right)
    \times \exp\left(-S_{\text{pert}}(\mu_{b}, \mu)\right) , 
    \label{g}
\end{align}
where $f_1^a(x,\mu)$ is the collinear PDF. $C_{g/a}$ is the perturbative coefficient function. The perturbative Sudakov factor is denoted by $S_\text{pert}$.
We can expand $C_{g/a}$ in powers of $\alpha_s$ as \cite{Boer:2020bbd}
\begin{align}
C_{g / a}\left(x, \mu_{b}\right)=\delta_{g a} \delta(1-x)+\sum_{k=1}^{\infty} C_{g / a}^{(k)}(x)\left(\frac{\alpha_{s}\left(\mu_{b}\right)}{\pi}\right)^{k}. \label{c1}
\end{align}
Within the next-to-leading logarithmic (NLL) approximation, only the leading-order (delta function) term of the coefficient function is considered, while the higher-order corrections of $\mathcal {O}(\alpha_s)$ are neglected.\\
Therefore, the convolution in Eq.~\eqref{g} can be simplified as
\begin{align}
\sum_{a = g, q, \bar{q}}
\left(C_{g/a} \otimes f_{1}^{a}\right)(x,\mu_b)
&=
\sum_{a = g, q, \bar{q}}
\int_x^1 \frac{\mathrm{d}x'}{x'}
\, C_{g/a}(x',\mu_b)\,
f_1^a\!\left(\frac{x}{x'},\mu_b\right) \nonumber\\
&=
\sum_{a = g, q, \bar{q}}
\int_x^1 \frac{\mathrm{d}x'}{x'}
\, \delta_{ga}\,\delta(1-x')\,
f_1^a\!\left(\frac{x}{x'},\mu_b\right)\nonumber\\
&= f_1^g(x,\mu_b). \label{conv}
\end{align}
Substituting Eq.~\eqref{conv} into Eq.~\eqref{g}, we obtain
\begin{align}
\tilde{f}_{1}^{g}\left(x, \boldsymbol{b}_{\sT}^2; \mu_b, \mu \right)
&= f_1^g(x,\mu_b)\,
\exp\!\left[-S_{\text{pert}}(\mu_b,\mu)\right].\label{eq:pert_tmd}
\end{align}

Let us now write the expression for the GSF, $f^{\perp g}_{1T}\left(x, \bm{q}_{\sT}^2; \mu \right)$, which is related to its derivative via a Fourier transformation~\cite{Aybat:2011ge}
\begin{align}
f^{\perp g}_{1T}\left(x, \bm{q}_{\sT}^2; \mu \right)
&= -\frac{1}{2\pi q_{\sT}}
\int_0^{\infty} \d b_\sT\, b_\sT\,
J_1(b_\sT q_{\sT})\,
\tilde{f}^{\prime\,\perp g}_{1T}\left(x, \bm{b}_{\sT}^2; \mu \right),\label{Sivers_Fourier}
\end{align}
where $J_1$ denotes the first-order Bessel function of the first kind.
The derivative of the GSF evolves in the same way as the unpolarized gluon TMD and is given by \cite{Echevarria:2014xaa,Aybat:2011ge}
\begin{align}
\tilde{f}^{\prime\,\perp g}_{1T}\left(x, \bm{b}_{\sT}^2; \mu_{b}, \mu \right)
&= \frac{b_\sT M_p}{2}\,T_{g,F}(x,x,\mu_{b})
\exp\!\left[-S_{\text{pert}}(\mu_{b},\mu)\right],
\label{eq:pert_sivers}
\end{align}
where $T_{g,F}(x,x,\mu_{b})$ is the Qiu--Sterman function, parametrized as \cite{Kouvaris:2006zy,Echevarria:2014xaa}
\begin{align}
T_{g,F}(x,x,\mu_{b}) = N_g(x)\,f_1^g(x,\mu_{b}).
\end{align}

Here, $N_g(x)$ is an $x$-dependent normalization function for the gluon Sivers distribution. At present, no experimental data are available to determine $N_g(x)$ directly. However, the corresponding normalization functions for the quark Sivers distributions have been extracted from SIDIS data through global fits and are given by \cite{Anselmino:2002pd,Anselmino:2008sga,Aybat:2011ge}
\begin{align}
N_f(x)=N_f\, x^{\alpha_f}(1-x)^{\beta_f}\,
\frac{(\alpha_f+\beta_f)^{(\alpha_f+\beta_f)}}{\alpha_f^{\alpha_f}\,\beta_f^{\beta_f}},\label{Ng}
\end{align}
where $N_f$, $\alpha_f$, and $\beta_f$, with $f$ denoting the quark flavour, are free parameters extracted from experimental data.

In this work, we consider two commonly used parametrizations for $N_g(x)$ \cite{Boer:2003tx}.
\begin{align}
\mathrm{TMD-a:}\quad N_g(x) = &\left[N_u(x) + N_d(x)\right]/2\,,\nonumber\\
\mathrm{TMD-b:}\quad N_g(x) =& N_d(x).
\label{TMD:ab}
\end{align}
We denote them by TMD-a and TMD-b, respectively.
\subsection{Perturbative Sudakov Factor}
The perturbative Sudakov factor for the gluon TMD is given by \cite{Maxia:2025zee,Catani:1988vd}
\begin{align}
S_{\text{pert}}(\mu_{b}, \mu) =
\int_{\mu_{b}^2}^{\mu^2} \frac{\mathrm{d}\eta^2}{\eta^2}
\left[
A_g\big(\alpha_s(\eta)\big) \log\left( \frac{\mu^2}{\eta^2} \right)
+ B_g\big(\alpha_s(\eta)\big)
\right],\label{pert_sudo}
\end{align}
where the coefficients $A_g$ and $B_g$ are the anomalous dimensions. They can be expanded perturbatively in powers of $\alpha_s$ as
\[
A_g\big(\alpha_s(\eta)\big) = \sum_{k=1}^\infty \left( \frac{\alpha_s(\eta)}{\pi} \right)^k A_g^{(k)}, 
\qquad
B_g\big(\alpha_s(\eta)\big) = \sum_{k=1}^\infty \left( \frac{\alpha_s(\eta)}{\pi} \right)^k B_g^{(k)}.
\]
Note that, in contrast to Eq.~(9) in Ref.~\cite{Maxia:2025zee}, we neglect both the additional soft contributions $B_\psi$ and $B_{ep}$, and hence do not consider the $B_{\text{CO}}$ term in Eq.~\eqref{pert_sudo}.\\
At NLL accuracy, $A_g$ at two-loop order and $B_g$ at one-loop order are given by \cite{Maxia:2025zee}
\begin{align*}
A_g^{(1)} &= \frac{C_A}{2}, \\
A_g^{(2)} &= \frac{C_A}{4} \left[
C_A \left( \frac{67}{18} - \frac{\pi^2}{6} \right)
- \frac{5}{9} n_f
\right], \\
B_g^{(1)} &= -\frac{C_A}{2}\,\frac{\beta_0}{6}, \qquad
\beta_0 = 11 - \frac{2}{3} n_f.
\end{align*}
By substituting the above coefficients and using the one-loop expansion of $\alpha_s$, 
the perturbative Sudakov factor at one-loop accuracy is given by \cite{Maxia:2025zee}
\begin{equation}
S_{\text{pert}}^{\text{1-loop}}(\mu_b, \mu)
= - \frac{4 A_g^{(1)}}{\beta_0} 
\left[
\log \left( \frac{\mu^2}{\mu_b^2} \right)
- \left( \frac{B_g^{(1)}}{A_g^{(1)}} + L_\mu \right)
\log \left( \frac{L_\mu}{L_{\mu_b}} \right)
\right].
\label{eq:S1loop}
\end{equation}
At NLL accuracy, it becomes
\begin{equation}
S_{\text{pert}}^{\text{NLL}}(\mu_b, \mu)
= S_{\text{pert}}^{\text{1-loop}}
- \frac{16 A_g^{(2)}}{\beta_0^2}
\left[
\log \left( \frac{L_\mu}{L_{\mu_b}} \right)
- \frac{1}{L_{\mu_b}} 
\log \left( \frac{\mu^2}{\mu_b^2} \right)
\right],
\label{eq:SNLL}
\end{equation}
where $L_\eta = \log \left( \eta^2 / \Lambda_{\text{QCD}}^2 \right)$.\\

The perturbative expressions in Eqs.~\eqref{eq:pert_tmd} and \eqref{eq:pert_sivers} are valid only in the small-$b_{\sT}$ region, $b_T \ll 1/\Lambda_{\mathrm{QCD}}$. However, to perform the integrals in Eqs.~\eqref{Fourier} and \eqref{Sivers_Fourier}, the integration must run from small to large values of $b_{\sT}$. In the large-$b_{\sT}$ region, perturbation theory fails because of large logarithmic contributions from higher-order terms. Therefore, the non-perturbative large-$b_{\sT}$ region cannot be described using perturbation theory alone and hence is not included in the integrals given by Eqs.~\eqref{Fourier} and \eqref{Sivers_Fourier}.

\subsection{$b_\ast$ Prescription}
To deal with the issue discussed above, we employ $b_\ast$ prescription \cite{Aybat:2011ge,Aybat:2011zv,Collins:2011zzd}, which allows us to use perturbation theory in the region where it is valid, while including non-perturbative effects separately, the modified impact parameter is defined as 
\[
b_\ast(b_\sT) = \frac{b_\sT}{\sqrt{1 + \left( \frac{b_\sT}{b_{\text{max}}} \right)^2 }}\, ,
\]
where $b_{\text{max}}$ is the upper cutoff, and we take $b_{\text{max}} =1.5~\text{GeV}^{-1}$.
For $b_T \ll b_{\text{max}}$, we have $b_\ast \approx b_T$, and the calculation remains perturbative. For $b_T \gg b_{\text{max}}$, the function $b_\ast$ saturates at the upper cutoff $b_{\text{max}}$, preventing the scale from entering the non-perturbative region. In addition, one must treat the small-$b_{\sT}$ region carefully. For small $b_{\sT}$, the scale $\mu_b = b_0/b_{\sT}$ can be larger than the hard scale $\mu$, leading to an unphysical enhancement of the Sudakov factor. To prevent this, we introduce a lower cutoff on $b_{\sT}$ through the parameter
\cite{Maxia:2025zee}
\[
b_{\min} = \frac{b_0}{\mu}.
\]
We then replace the scale $\mu_b$ by a regulated scale
\begin{equation}
\mu_{b_\ast}' = \frac{b_0}{\sqrt{b_\ast^2 + b_{\min}^2}},
\end{equation}
which ensures the scale associated with $b_{\sT}$ always remains within a physically meaningful perturbative range.
The non-perturbative effects that dominate at large $b_T$ are incorporated separately through the factor $e^{-S_{\text{NP}}}$, which restores the correct $b_T$-dependence beyond $b_{\text{max}}$.

The modified unpolarized gluon TMD and gluon Sivers function are then written as 
\begin{align} 
\tilde{f}_{1}^{g}\left(x, \bm{b}_{\sT}^2 ; \mu_{b_\ast}',\mu\right) 
&=f_1^g(x,\mu_{b_\ast}')\,
\exp\!\left[-S_{\text{pert}}(\mu_{b_\ast}',\mu)\right]\,e^{-S_{\text{NP}}}\,, 
\label{final_tmd}
\end{align} 
and
\begin{align}
\tilde{f}^{\prime\,\perp g}_{1T}\left(x, \bm{b}_{\sT}^2; \mu_{b_\ast}', \mu \right)
&= \frac{b_\sT M_p}{2}\,T_{g,F}(x,x,\mu_{b_\ast}')
\exp\!\left[-S_{\text{pert}}(\mu_{b_\ast}',\mu)\right]\,e^{-S_{\text{NP}}}\,,
\label{final_sivers}
\end{align}
where Eqs.~\eqref{final_tmd} and \eqref{final_sivers} are obtained from Eqs.~\eqref{eq:pert_tmd} and \eqref{eq:pert_sivers}, respectively, by replacing $\mu_b \to \mu_{b_\ast}'$ and including the factor $e^{-S_{\text{NP}}}$.
$S_{\text{NP}}$ denotes the non-perturbative Sudakov factor which parametrizes the large-$b_T$ behaviour of the TMD.
\subsection{Non-perturbative Sudakov Factor}
The non-perturbative Sudakov factor for the unpolarized TMD is parametrized as \cite{Echevarria:2014xaa}
\begin{align}
S_{\text{NP}}^{\text{pdf}}(b_\sT,\mu) &= b_{\sT}^2 \left( g_1^{\text{pdf}} + \frac{g_2}{2} \log \frac{\mu}{Q_0} \right),
\end{align}
while for the Sivers function, we use the parametrization \cite{Echevarria:2014xaa}
\begin{align}
S_{\text{NP}}^{\text{Sivers}}(b_\sT,\mu) &= b_{\sT}^2 \left( g_1^{\text{Sivers}} + \frac{g_2}{2} \log \frac{\mu}{Q_0} \right),
\end{align}
where $\mu=\sqrt{M_\psi^2+Q^2}$.
The coefficients correspond to the average of intrinsic transverse momentum squared at the scale $Q_0$:
\[
g_1^{\text{pdf}} = \frac{\langle k_\perp^2 \rangle_{Q_0}}{4}, 
\qquad 
g_1^{\text{Sivers}} = \frac{\langle k_{s\perp}^2 \rangle_{Q_0}}{4}.
\]
In Ref. \cite{Echevarria:2014xaa},  the best fit parameters were extracted by fitting the single spin asymmetry SIDIS data of pion, kaon, and charged-hadron production at JLab, HERMES, and COMPASS. The best fit parameters used in this work are 
\begin{align}
N_u &= 0.106, \qquad & N_d &=-0.163, \qquad & N_s &= 0.103, \nonumber \\
N_{\bar{u}} &= -0.012, \qquad & N_{\bar{d}} &=-0.105, \qquad & N_{\bar{s}} &= -1.000, \nonumber \\
\alpha_u &= 1.051, \qquad & \alpha_d &=1.552, \qquad & \alpha_{{\text{sea}}} &= 0.851, \qquad & \beta & =4.857
\end{align}
and
\begin{equation}
\langle k_\perp^2 \rangle_{Q_0} = 0.38~\text{GeV}^2, \quad
\langle k_{s\perp}^2 \rangle_{Q_0} = 0.282~\text{GeV}^2, \quad
g_2 = 0.16~\text{GeV}^2.
\end{equation}
All the parameters were fitted at the initial scale $Q_0 = \sqrt{2.4}\,\text{GeV}$.\\
By substituting Eqs.~\eqref{Fourier} and \eqref{Sivers_Fourier} into Eq.~\eqref{ratio_gamma} and Eq.~\eqref{ratio_jet}, we estimate the asymmetries corresponding to the $J/\psi+\gamma$ and $J/\psi+\mathrm{jet}$ processes, respectively.
\newpage
\section{Numerical Results}
\label{sec:results}
\subsection{Unpolarized cross section}

\begin{figure}[H]
    \centering
    
    \begin{subfigure}{0.48\textwidth}
        \centering
        \includegraphics[width=\linewidth]{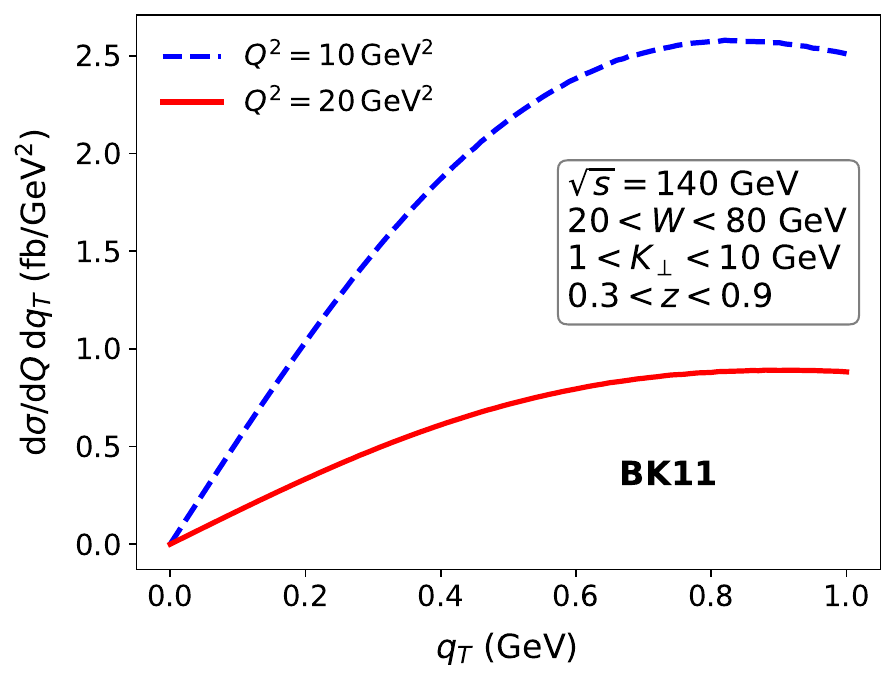}
        \caption{}
    \end{subfigure}
    \hfill
    \begin{subfigure}{0.48\textwidth}
        \centering
        \includegraphics[width=\linewidth]{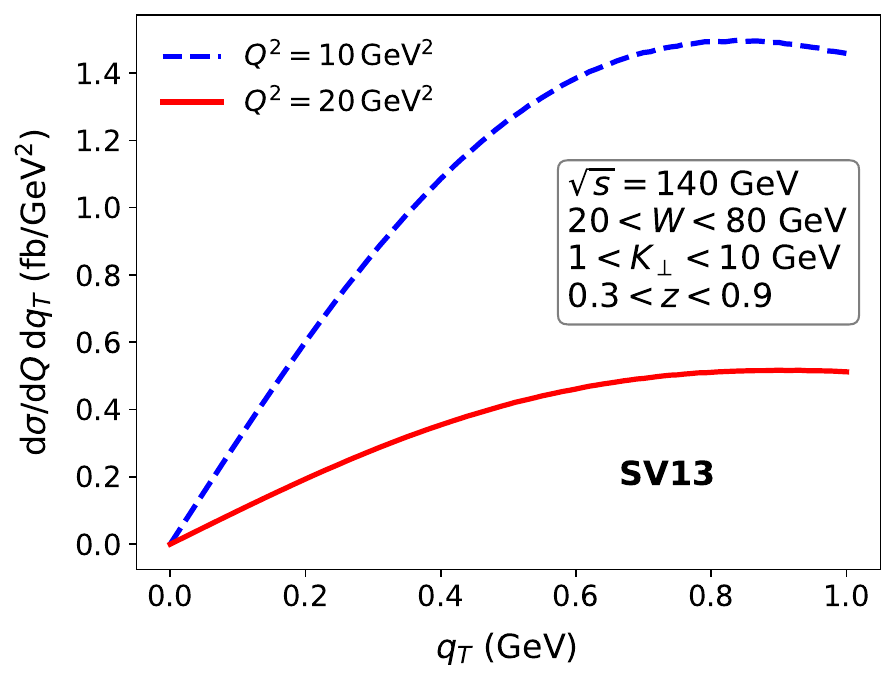}
        \caption{}
    \end{subfigure}
    
    \vspace{0.4cm}
    
    \begin{subfigure}{0.48\textwidth}
        \centering
        \includegraphics[width=\linewidth]{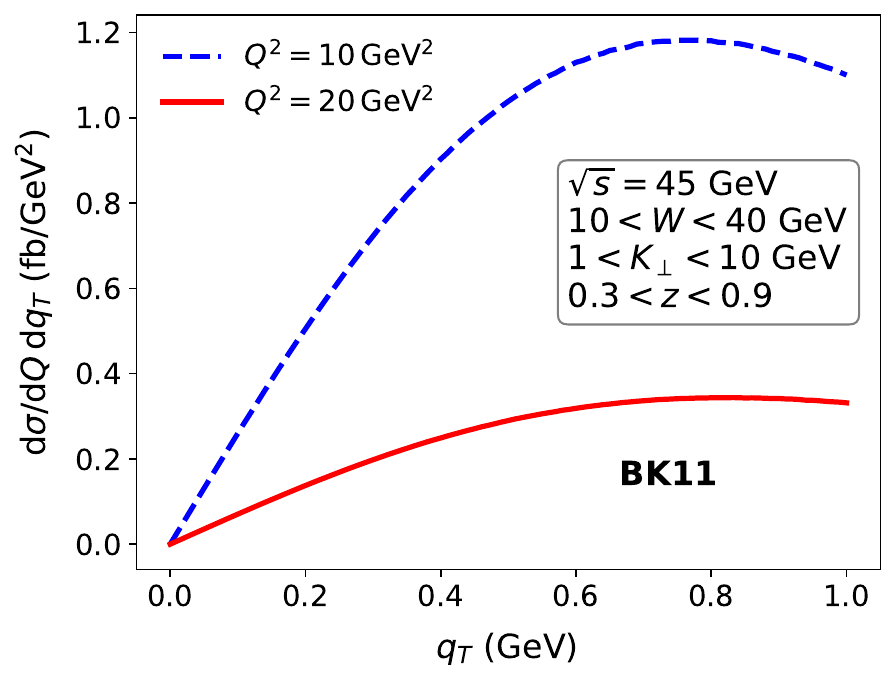}
        \caption{}
    \end{subfigure}
    \hfill
    \begin{subfigure}{0.48\textwidth}
        \centering
        \includegraphics[width=\linewidth]{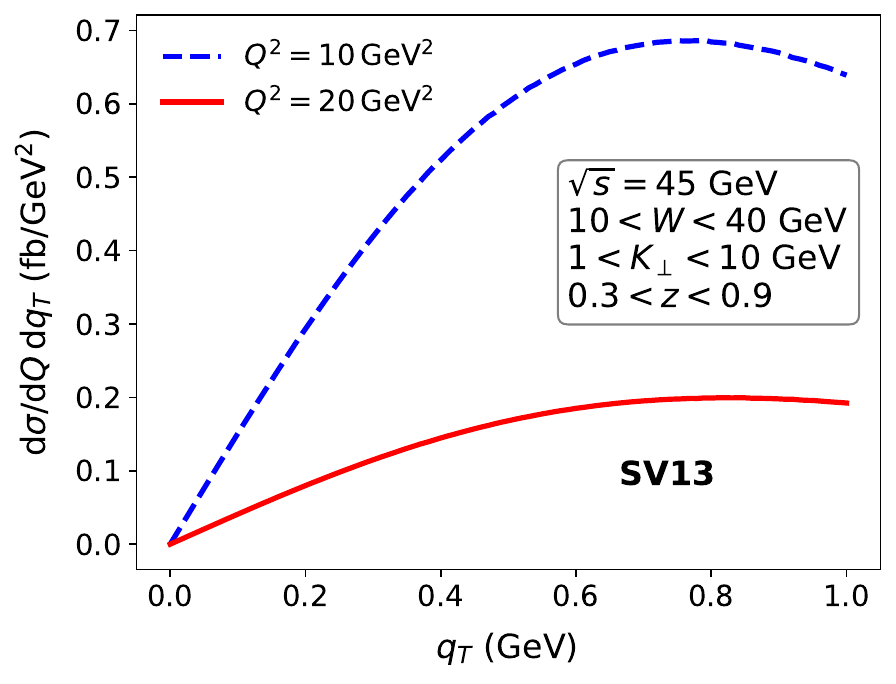}
        \caption{}
    \end{subfigure}

    \caption{\label{fig:2}
    Unpolarized differential cross section in $e+p \rightarrow e+J/\psi+\gamma +X$ 
    process as a function of $q_\sT$ for $Q^2=10~\mathrm{GeV}^2$ and $Q^2=20~\mathrm{GeV}^2$  at the EIC kinematics  $\sqrt{s}=140$ GeV (top panels) and  $\sqrt{s}=45$ GeV (bottom panels). The $z$ and $K_\perp$ are integrated over the ranges  $0.3 < z < 0.9$ and $1 < K_\perp < 10~\mathrm{GeV}$, respectively. The integration range of $W$ is from $10 < W < 40$ GeV for $\sqrt{s}=45$ GeV, while $20 < W < 80$ GeV for $\sqrt{s}=140$ GeV. Panels (a) and (c) correspond to the BK11 LDME set, while panels (b) and (d) correspond to the SV13 LDME set.}
\end{figure}

\begin{figure}[H]
    \centering
    
    \begin{subfigure}{0.48\textwidth}
        \centering
        \includegraphics[width=\linewidth]{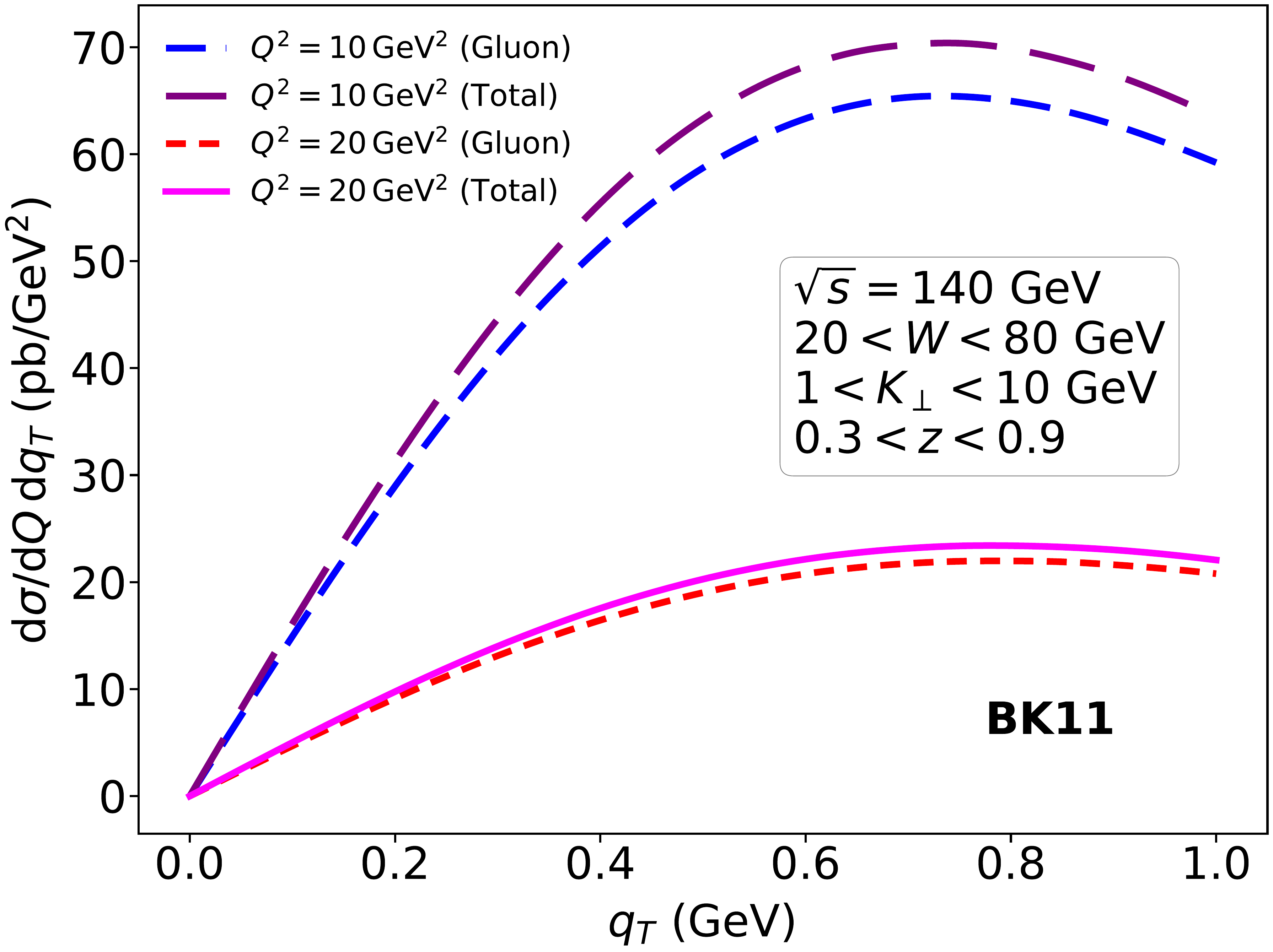}
        \caption{}
    \end{subfigure}
    \hfill
    \begin{subfigure}{0.48\textwidth}
        \centering
        \includegraphics[width=\linewidth]{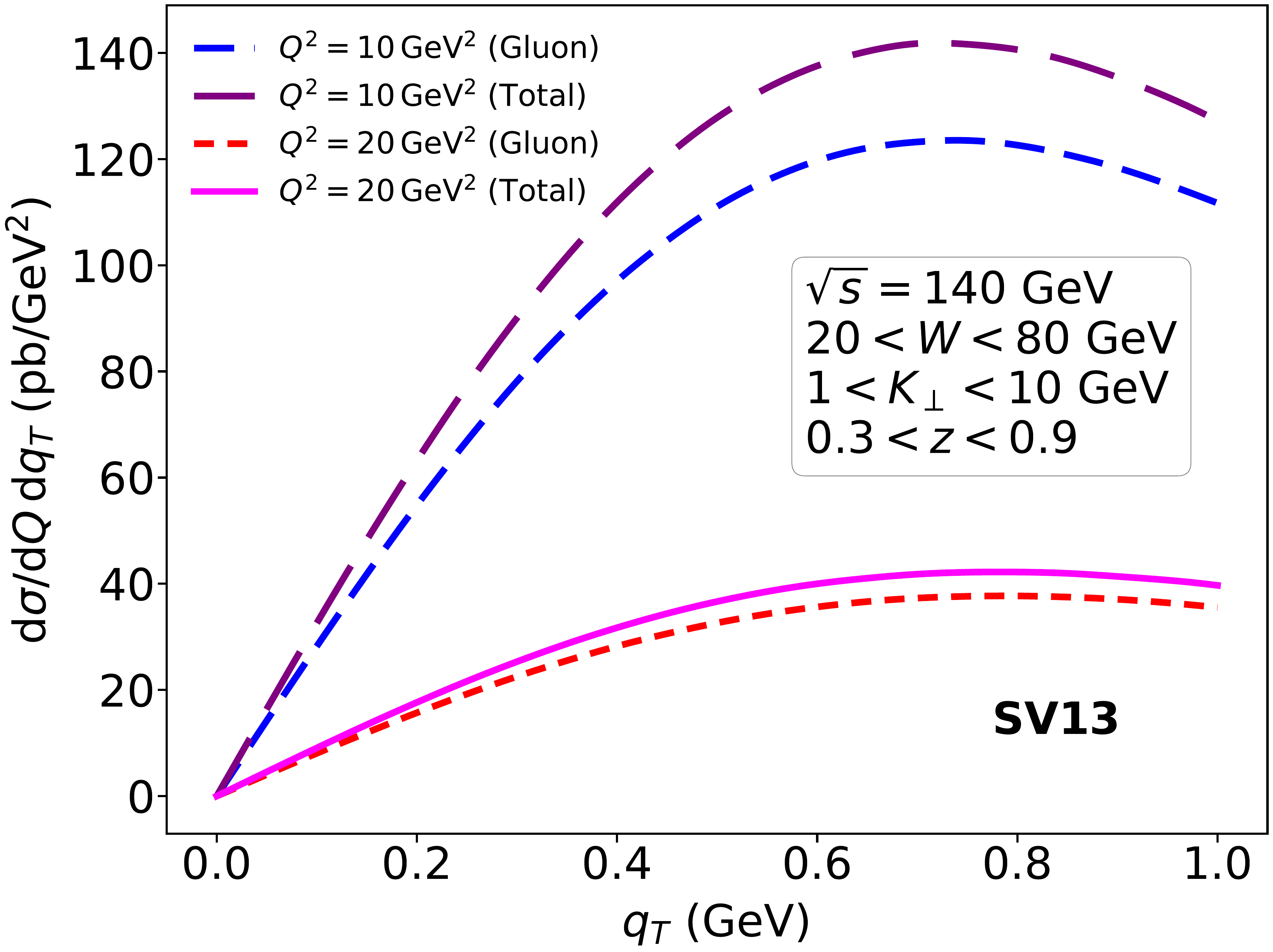}
        \caption{}
    \end{subfigure}
    
    \vspace{0.4cm}
    
    \begin{subfigure}{0.48\textwidth}
        \centering
        \includegraphics[width=\linewidth]{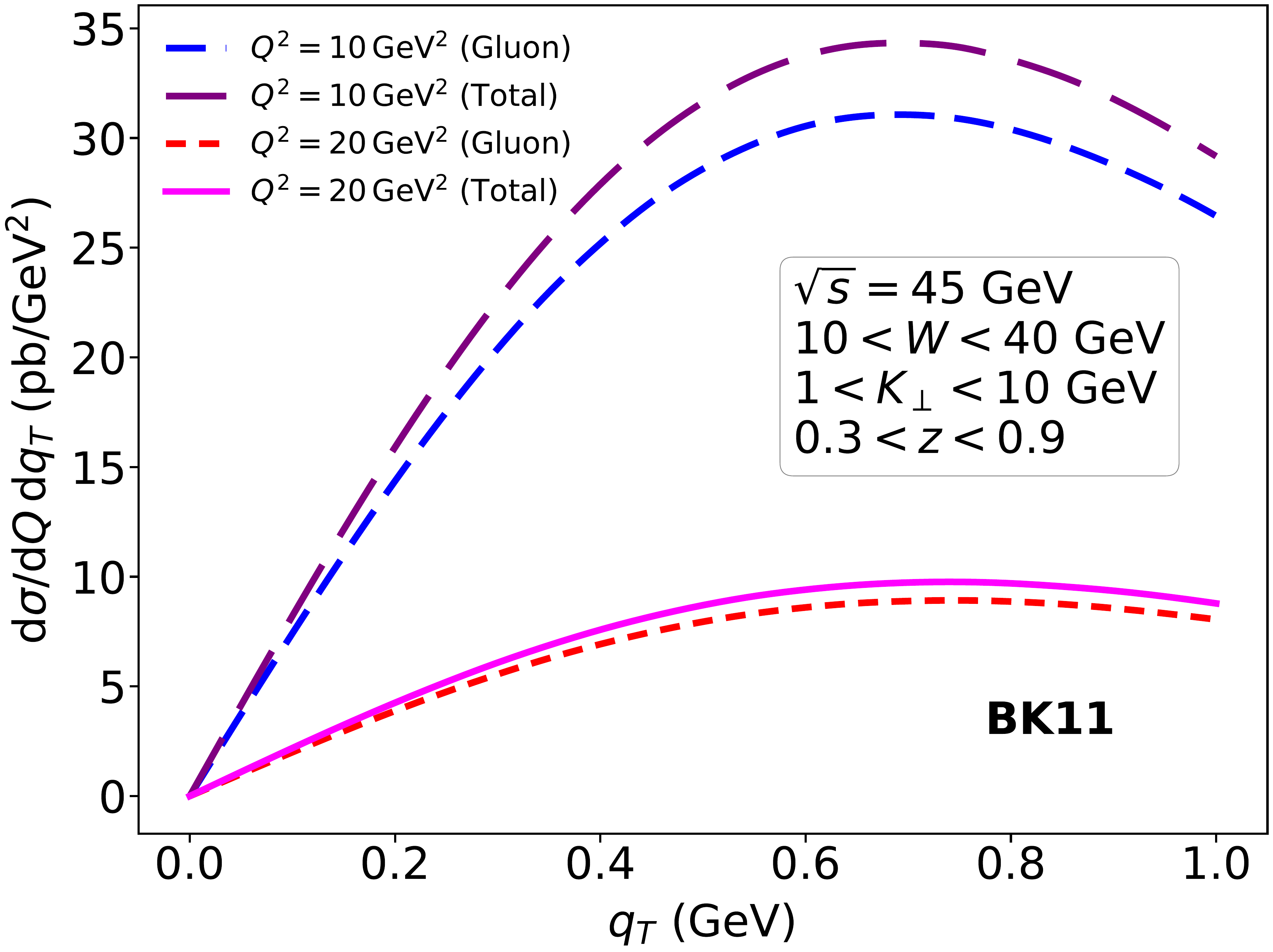}
        \caption{}
    \end{subfigure}
    \hfill
    \begin{subfigure}{0.48\textwidth}
        \centering
        \includegraphics[width=\linewidth]{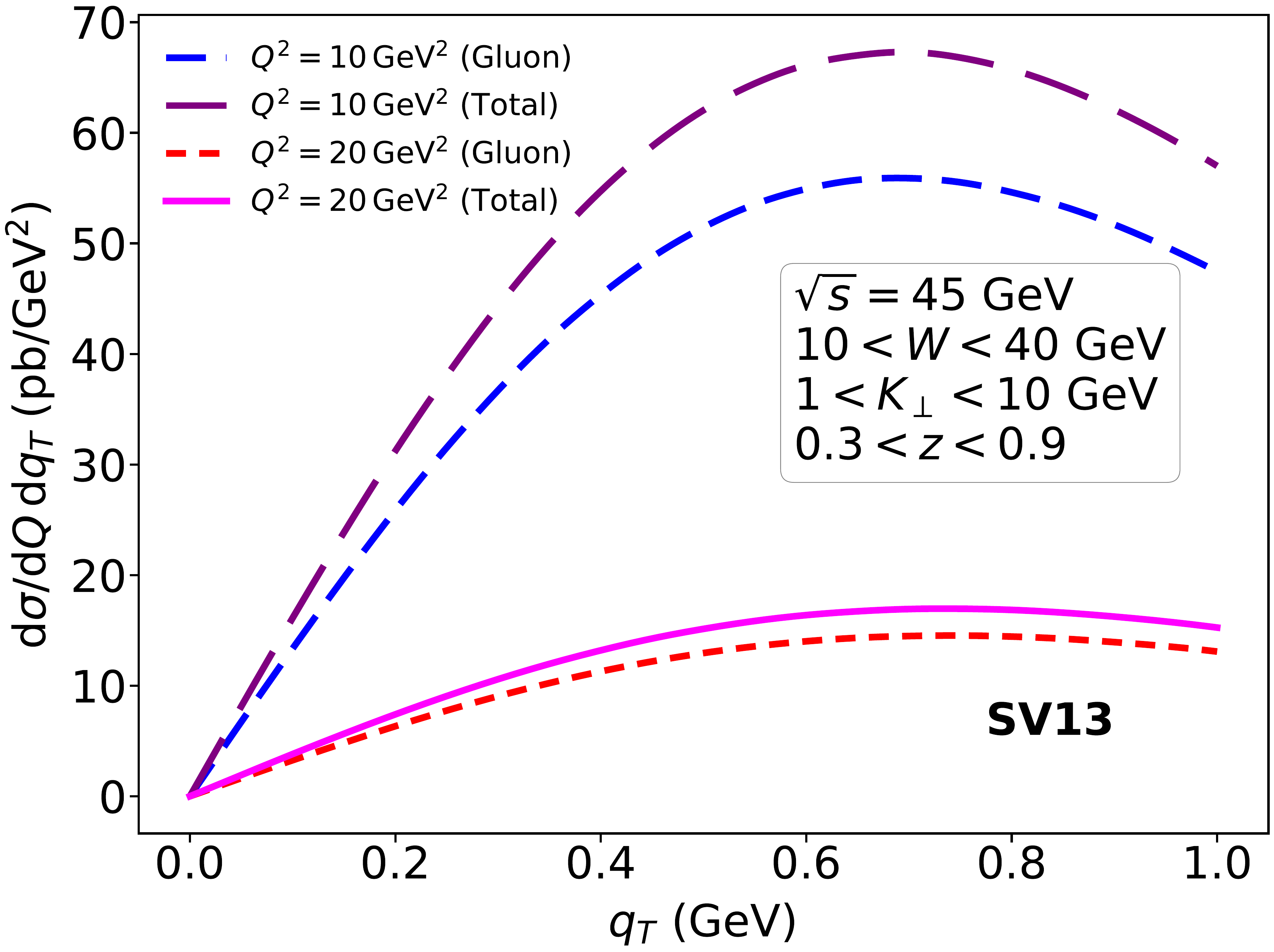}
        \caption{}
    \end{subfigure}

    \caption{\label{fig:3}
    Unpolarized differential cross section in $e+p \rightarrow e+J/\psi+\mathrm{jet} +X$ 
    process as a function of $q_\sT$ for $Q^2=10~\mathrm{GeV}^2$ and $Q^2=20~\mathrm{GeV}^2$ at the EIC kinematics  $\sqrt{s}=140$ GeV (top panels) and  $\sqrt{s}=45$ GeV (bottom panels). The $z$ and $K_\perp$ are integrated over the ranges  $0.3 < z < 0.9$ and $1 < K_\perp < 10~\mathrm{GeV}$, respectively. The integration range of $W$ is from $10 < W < 40$ GeV for $\sqrt{s}=45$ GeV, while $20 < W < 80$ GeV for $\sqrt{s}=140$ GeV. Panels (a) and (c) correspond to the BK11 LDME set, while panels (b) and (d) correspond to the SV13 LDME set.}
\end{figure}

The numerical results are presented in this section. In Figs.~\ref{fig:2} and \ref{fig:3}, we show the unpolarized differential cross section as a function of $q_{\sT}$, the sum of the transverse momentum of $J/\psi$ and photon(jet), for two cm energies 45 GeV and 140 GeV at EIC kinematics. As we consider the kinematical configuration where the $J/\psi$ and photon(jet) are almost back-to-back in the transverse plane, the transverse momentum $\bm K_\perp$, approximately equal to that of either $J/\psi$ or the photon (jet), is very large compared to $\bm q_\sT$. We use TMD factorization in this kinematics to describe the cross section.
The numerical results are shown for two different photon virtualities, $Q^2 = 10~\text{GeV}^2$ and $20~\text{GeV}^2$. The unpolarized cross section is obtained by integrating the kinematical variables in the following ranges: $0.3 < z < 0.9$ and $1 < K_\perp < 10~\mathrm{GeV}$. The lower cut $z>0.3$ is imposed to suppress the resolved-photon contributions relevant at low $z$, while the upper cut $z<0.9$ excludes contributions from diffractive processes through Pomeron exchange and avoids infrared divergences near $z\approx1$. The invariant mass of photon-proton system $W$ is integrated from $10 < W < 40~\text{GeV}$ for cm energy 45~\text{GeV}, while $20 < W < 80~\text{GeV}$ for cm energy 140~\text{GeV}. We present the results in the kinematical region that will be accessed at the upcoming EIC. The NRQCD framework is used for quarkonium production, wherein the initial heavy quark pair can be formed both in color singlet and color octet states. Two sets of LDMEs are considered, namely, BK11~\cite{Butenschoen:2011yh} and SV13~\cite{Sharma:2012dy}, which were extracted at the scale $M_\psi$. While the collinear PDFs are computed at $\mu_b$ using the LHAPDF grid of the CT18NLO~\cite{Hou:2019efy} PDF set.
In Figs.~\ref{fig:2} and~\ref{fig:3}, panels (a) and (b) correspond to $\sqrt{s}=140~\text{GeV}$, while panels (c) and (d) correspond to $\sqrt{s}=45~\text{GeV}$. The BK11 LDME set is used in panels (a) and (c), whereas the SV13 LDME set is used in panels (b) and (d).

\begin{table}[h!]
\centering
\resizebox{\textwidth}{!}{%
\begin{tabular}{|c|c|c|c|c|}
\hline
LDME set
 & $\langle 0| \mathcal{O}^{J/\psi}_{8} ({}^{1}S_{0}) |0 \rangle \;(10^{-2}\,\text{GeV}^3)$ 
 & $\langle 0| \mathcal{O}^{J/\psi}_{8} ({}^{3}S_{1}) |0 \rangle \;(10^{-3}\,\text{GeV}^3)$
 & $\langle 0| \mathcal{O}^{J/\psi}_{1} ({}^{3}S_{1}) |0 \rangle \;(\text{GeV}^3)$
 & $\langle 0| \mathcal{O}^{J/\psi}_{8} ({}^{3}P_{0}) |0 \rangle \;(10^{-2}\,\text{GeV}^5)$ \\
\hline

BK11~\cite{Butenschoen:2011yh} 
& $4.97\pm 0.44$ 
& $2.24\pm 0.59$ 
& $1.32$ 
& $-1.61\pm 0.20$ \\

\hline

SV13~\cite{Sharma:2012dy} 
& $1.80\pm 0.87$ 
& $1.30\pm 1.30$ 
& $1.20$ 
& $3.5280\pm 1.7052$ \\

\hline
\end{tabular}
}
\caption{Numerical values of the LDMEs for the BK11 and SV13 sets.}
\end{table}
In Fig.~\ref{fig:2}, the unpolarized cross section for the production of $J/\psi+\gamma$ is shown as a function of $q_\sT$. Since only the CO $^{3}S_{1}$ gluon state contributes at leading order, we show only the gluon contribution in the $J/\psi+\gamma$ production. Moreover, the LDME value of this state in the BK11 set is roughly 1.73 times larger than that of SV13. As a result, the cross section using the BK11 set is approximately 1.73 times larger than the SV13 set. The unpolarized cross section for $J/\psi + \text{jet}$ production is shown in Fig.~\ref{fig:3}. Here, since both the gluon and quark\,(anti-quark) channels contribute at leading order, we show separately the gluon contribution and the total contribution obtained by including both gluon and quark\,(anti-quark) channels. Whereas in $J/\psi + \text{jet}$ production, both the CS and CO states contribute. In contrast to the SV13 set, the cross section has been suppressed due to the negative LDME value of the CO  $^{3}P_{0}$ state in the BK11 set.  In general,   the SV13's set cross section is twice as large as the BK11 set. Additionally, the CS state contribution dominates over the CO states in  $J/\psi + \text{jet}$ production. Interestingly, we found that the contribution from the quark channel is much smaller than that from the gluon channel. This advocates that  $J/\psi + \gamma~\text{(jet)}$ production is an ideal channel to probe the unknown gluon TMDs.
By comparing Figs.~\ref{fig:2} and~\ref{fig:3}, we found that the $J/\psi + \text{jet}$ cross section is significantly larger than the $J/\psi + \gamma$ cross section. As $Q^2$ increases, the cross section decreases due to the suppression from the virtual photon propagator, which scales as $1/Q^4$. In contrast, the cross section increases with increasing $\sqrt{s}$, since higher cm energy probes a smaller longitudinal momentum fraction $x$, where the gluon density is larger.

The $q_\sT$-dependence of the cross section can be qualitatively understood from the following expression.
\begin{align}
    \frac{\d\sigma}{\d q_\sT} \propto q_\sT \int_0^\infty \d b_\sT\,b_\sT\,J_0(b_\sT q_\sT)\,f_1^g(x,\mu_{b_\ast}')\,e^{-S(b_\sT,Q)}|H|^2 \,,
\end{align}
where the momentum fraction $x$, Sudakov factor $S$ and hard part $H$ do not depend on $q_\sT$.
At small $q_\sT$, the cross section shows a linear rise due to the explicit $q_\sT$ prefactor. In this region, the Bessel function $J_0(b_\sT q_\sT)$ is close to unity. As $q_T$ increases, the Bessel function starts to oscillate. This creates a competition with the growth from the $q_\sT$ factor and leads to a maximum. At larger $q_\sT$, the oscillatory nature of $J_0(b_\sT q_\sT)$ causes positive and negative contributions in the $b_T$ integral to partially cancel, and since this cancellation becomes stronger as $q_\sT$ increases, the value of the integral decreases.

\subsection{Sivers asymmetry}
The Sivers asymmetry for the production of $J/\psi+\gamma$ and $J/\psi+\mathrm{jet}$ is estimated at fixed values of  $Q^2=10~\text{GeV}^2$, $K_\perp=10~\text{GeV}$, $q_\sT=1~\mathrm{GeV}$, $z=0.4$, and $y=0.1$ (for 140 GeV) and $y=0.8$ (for 45 GeV). These values are chosen particularly to maximize the asymmetry. Unfortunately, the GSF has not been extracted within the TMD evolution approach; hence, we use two parametrizations for the GSF as defined in Eq.~\eqref{TMD:ab}. For the TMD-a parametrization, the $N_g(x)$ of the GSF is defined by averaging the $N_u(x)$ and $N_{d}(x)$ of up-quark and down-quark Sivers functions, respectively. Moreover, the best fit value $N_d(x)$ for the down-quark Sivers function is negative, while $N_u(x)$ is positive for the up-quark Sivers function. This leads to a smaller value of $N_g(x)$  for the GSF in the TMD-a parametrization. As a result, the Sivers asymmetry is maximum for the TMD-b parametrization compared to the TMD-a parametrization in all the figures.

\begin{figure}[H]
    \centering
    \begin{subfigure}{0.48\textwidth}
        \centering
        \includegraphics[width=\linewidth]{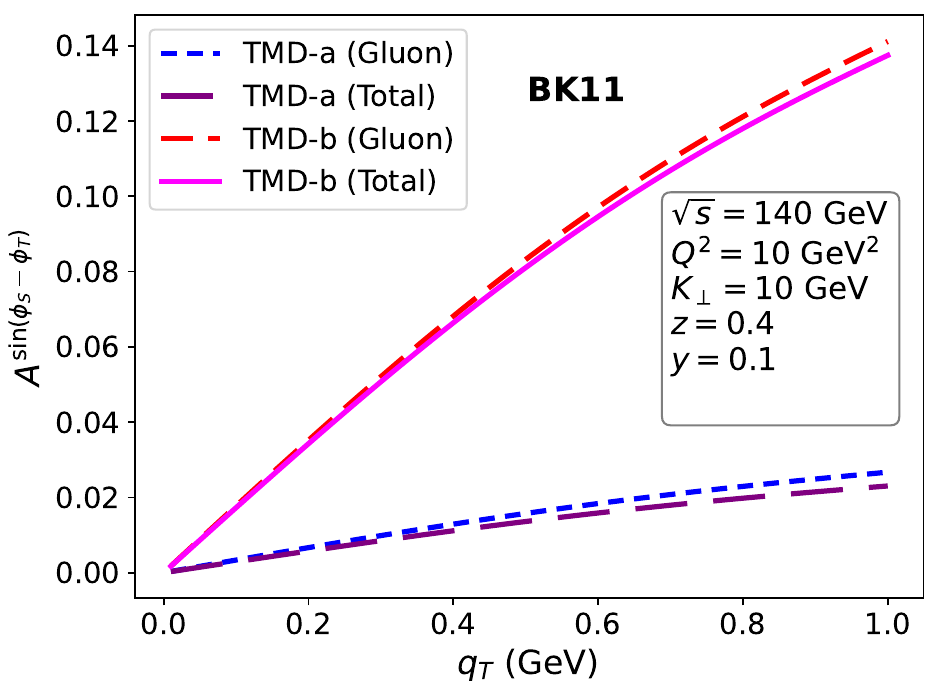}
        \caption{}
    \end{subfigure}
    \hfill
    \begin{subfigure}{0.48\textwidth}
        \centering
        \includegraphics[width=\linewidth]{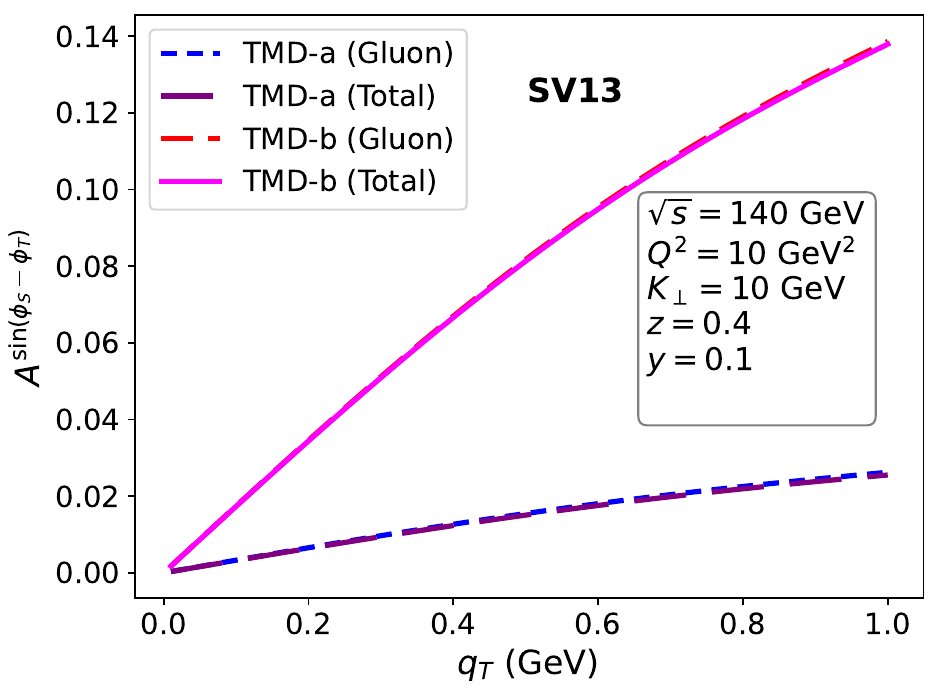}
        \caption{}
    \end{subfigure}
    
    \vspace{0.4cm}
    
    \begin{subfigure}{0.48\textwidth}
        \centering
        \includegraphics[width=\linewidth]{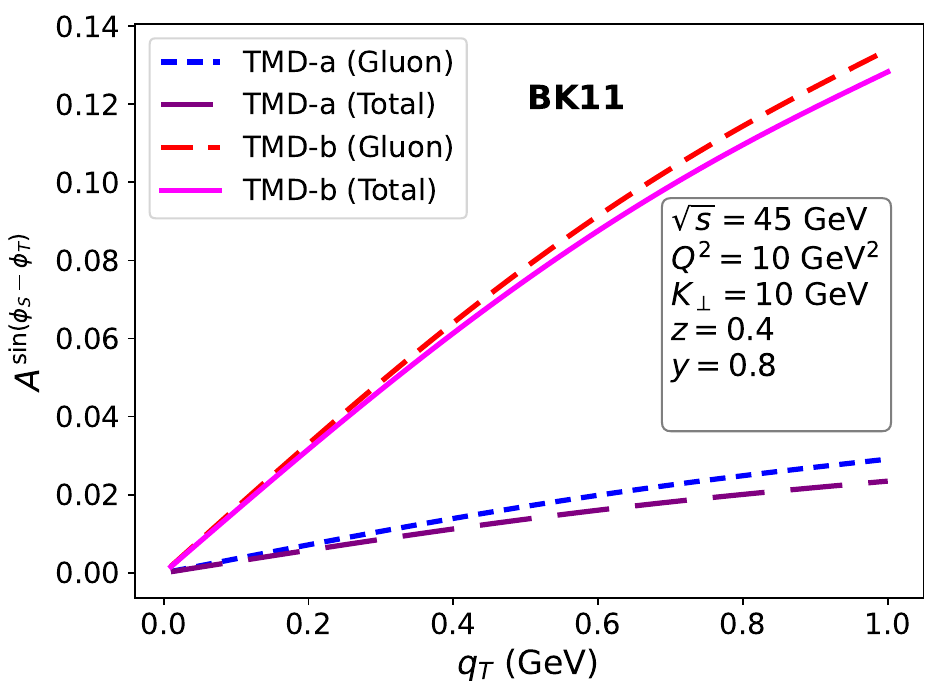}
        \caption{}
    \end{subfigure}
    \hfill
    \begin{subfigure}{0.48\textwidth}
        \centering
        \includegraphics[width=\linewidth]{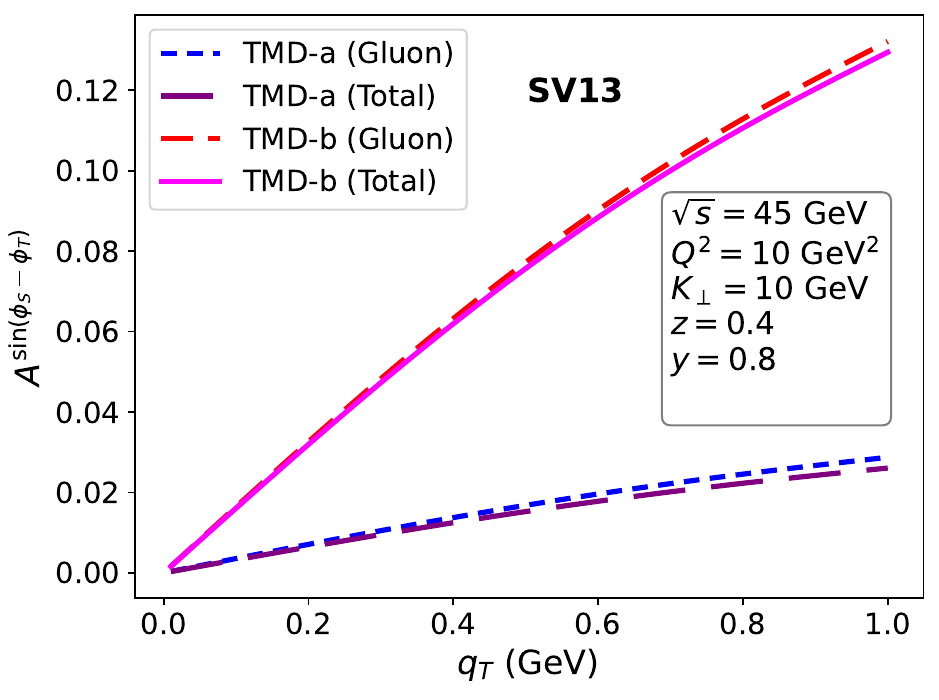}
        \caption{}
    \end{subfigure}
    \caption{\label{fig:5}
    Sivers asymmetry in $e+p^\uparrow\rightarrow e+J/\psi+\mathrm{jet} +X$ process as a function of $q_T$ at the EIC kinematics $\sqrt{s}=140$ GeV (top panels) and  $\sqrt{s}=45$ GeV (bottom panels). The asymmetry is estimated at fixed values of $Q^2=10~\text{GeV}^2$, $K_\perp=10~\text{GeV}$, $z=0.4$, and $y=0.1$ (for 140 GeV) and $y=0.8$ (for 45 GeV). Panels (a) and (c) correspond to the BK11 LDME set, while panels (b) and (d) correspond to the SV13 LDME set.}
\end{figure}

\begin{figure}[H]
    \centering
     \begin{subfigure}{0.48\textwidth}
        \centering
        \includegraphics[width=\linewidth]{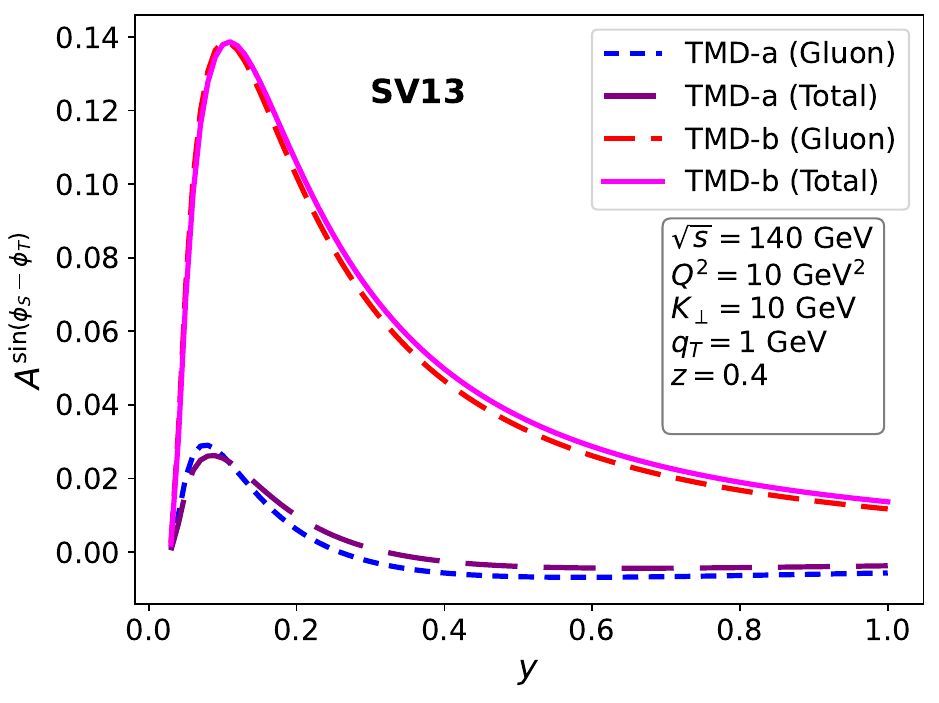}
        \caption{}
    \end{subfigure}
    \hfill
    \begin{subfigure}{0.48\textwidth}
        \centering
        \includegraphics[width=\linewidth]{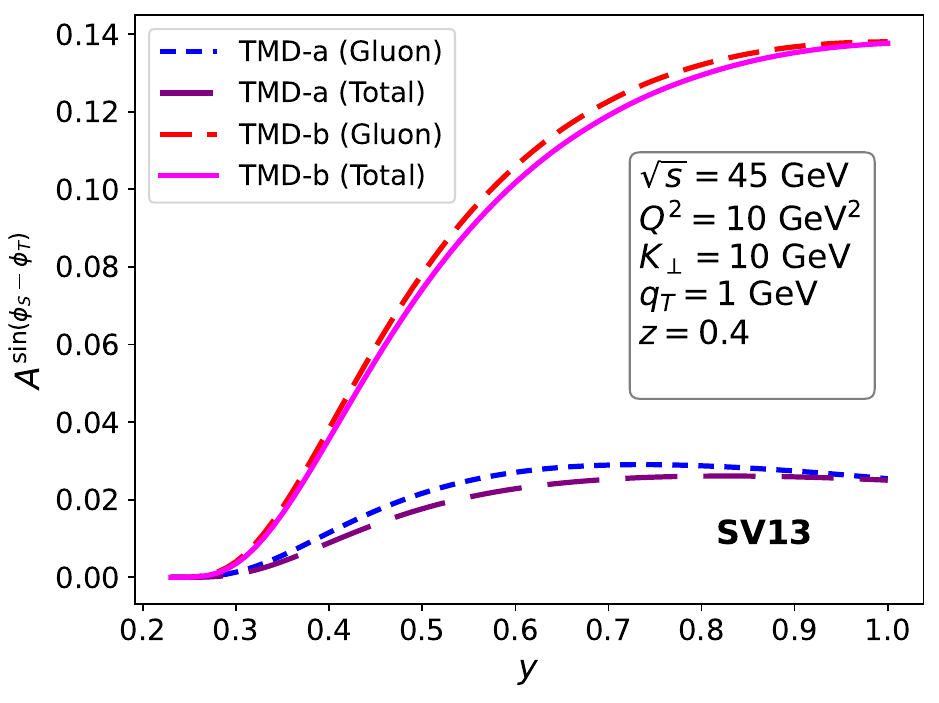}
        \caption{}
    \end{subfigure}
    \caption{\label{fig:7}
    Sivers asymmetry in $e+p^\uparrow\rightarrow e+J/\psi+\mathrm{jet} +X$ process as a function of $y$ at the EIC kinematics $\sqrt{s}=140$ GeV (left) and  $\sqrt{s}=45$ GeV (right). The asymmetry is estimated at fixed values of $Q^2=10~\text{GeV}^2$, $K_\perp=10~\text{GeV}$, $q_T=1~\text{GeV}$, and $z=0.4$, using the SV13 LDME set.}
\end{figure}

\begin{figure}[H]
    \centering
        \begin{subfigure}{0.48\textwidth}
        \centering
        \includegraphics[width=\linewidth]{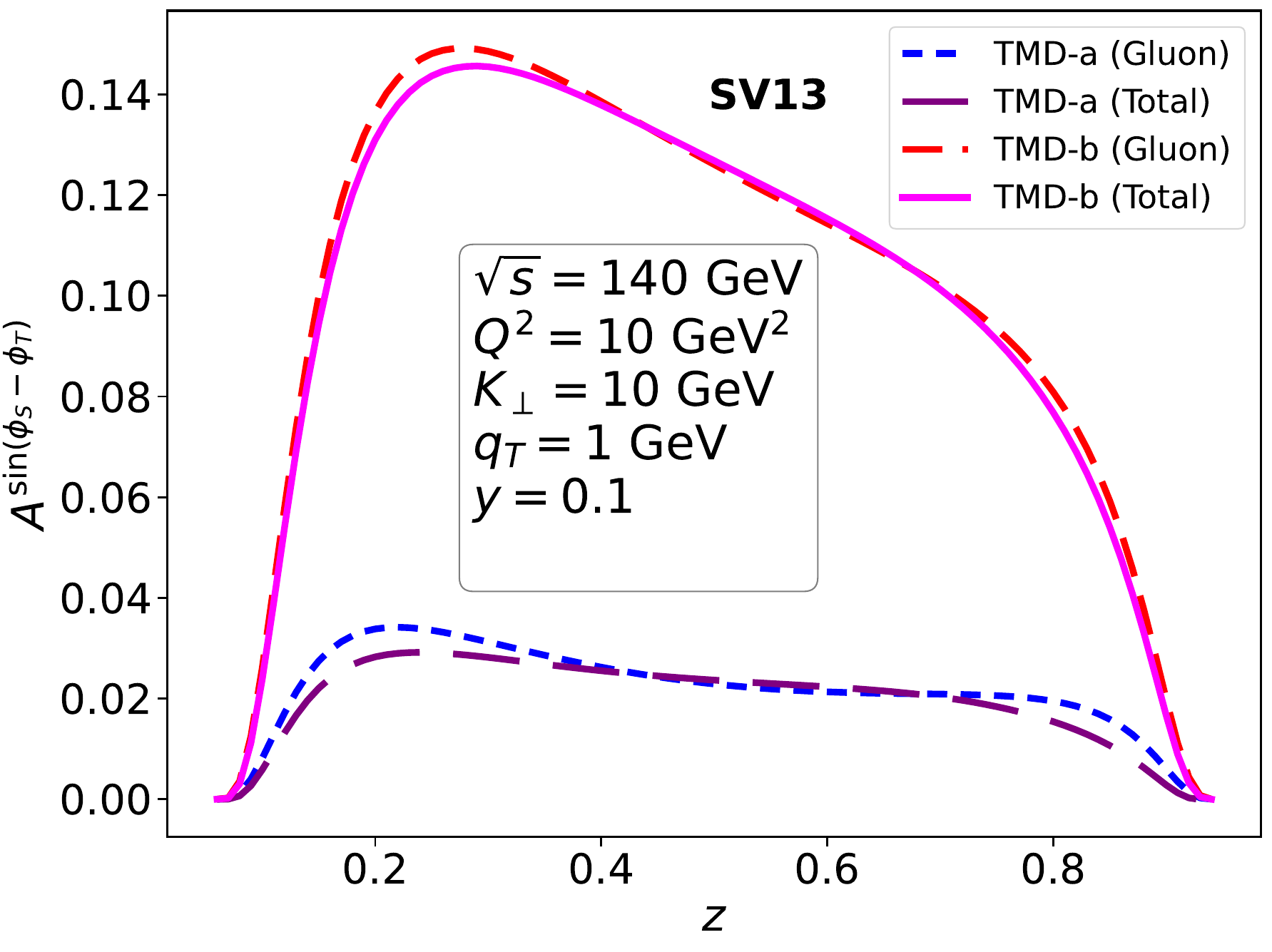}
        \caption{}
    \end{subfigure}
    \hfill
    \begin{subfigure}{0.48\textwidth}
        \centering
        \includegraphics[width=\linewidth]{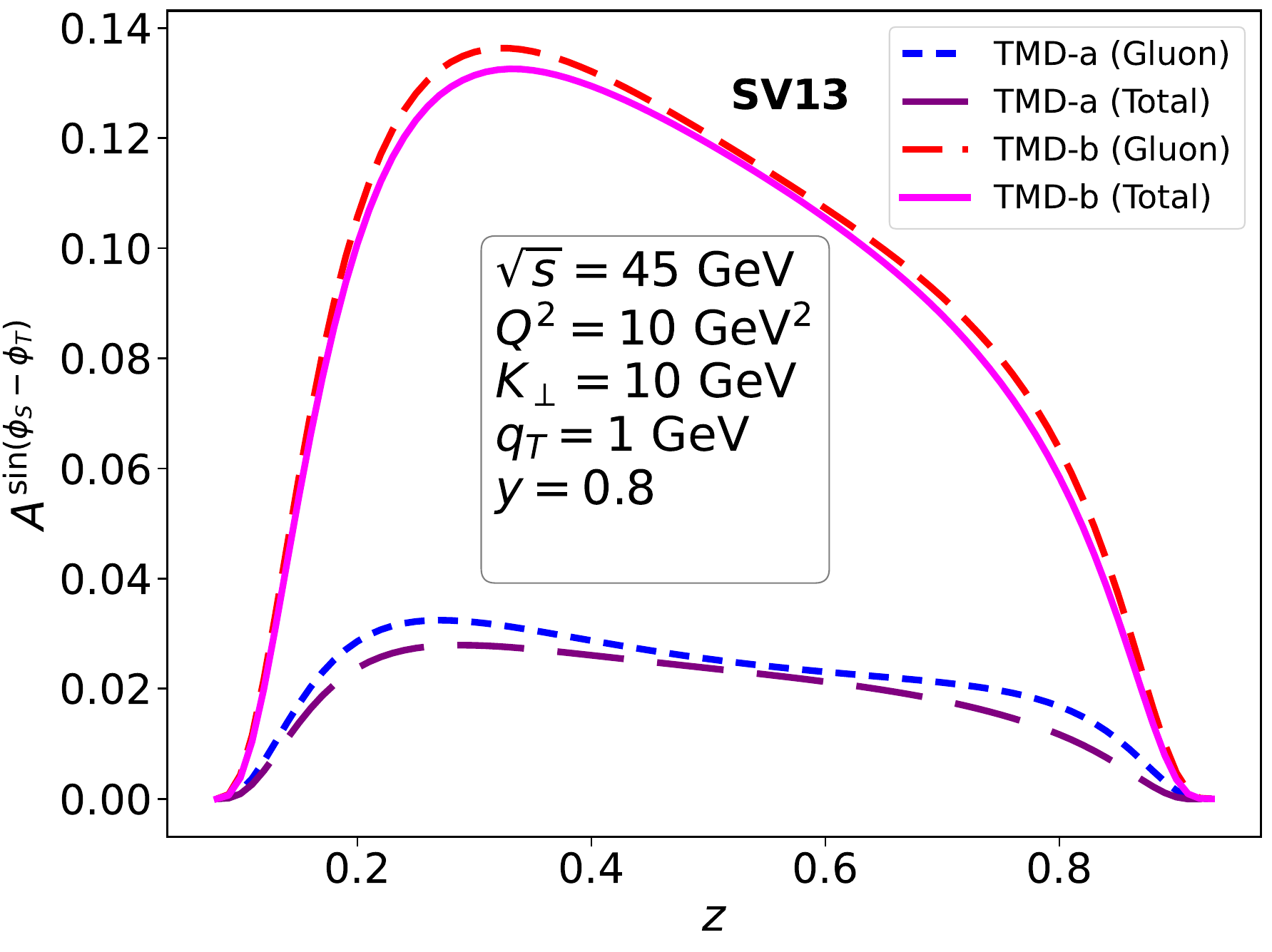}
        \caption{}
    \end{subfigure}
    \caption{\label{fig:9}
    Sivers asymmetry in $e+p^\uparrow\rightarrow e+J/\psi+\mathrm{jet} +X$ process as a function of $z$ at the EIC kinematics $\sqrt{s}=140$ GeV (left) and  $\sqrt{s}=45$ GeV (right). The asymmetry is estimated at fixed values of $Q^2=10~\text{GeV}^2$, $K_\perp=10~\text{GeV}$, $q_T=1~\text{GeV}$, and $y=0.1$ (for 140 GeV) and $y=0.8$ (for 45 GeV), using the SV13 LDME set.}
\end{figure}

Moreover, the estimated Sivers asymmetry for $J/\psi + \gamma$ production is found to be independent of the choice of LDME set, due to cancellation of the amplitude modulation factor $A_0^g$ as given in Eq.~\eqref{ratio_gamma}. The TMD-b parametrization predicts about  15$\%$ asymmetry, while the TMD-a estimates about 3$\%$ in all the figures of $J/\psi + \gamma$ production. For $J/\psi+\mathrm{jet}$ production, both gluon and quark (anti-quark) channels contribute to the formation of quarkonium. However, from Figs.~\ref{fig:5}, \ref{fig:7} and \ref{fig:9}, it is observed that the contribution from the quark (anti-quark) channel to the asymmetry is very small compared to that from the gluon channel. Due to this fact, the asymmetry does not depend strongly on the choice of LDME set even for $J/\psi + \mathrm{jet}$ production. Therefore, we present the asymmetry as a function of $q_\sT$ for both LDME sets only in Fig.~\ref{fig:5}, while the asymmetry as a function of $y$ and $z$ is shown only for the SV13 LDME set in Figs.~\ref{fig:7} and \ref{fig:9}, respectively. The predicted asymmetry is approximately 13$\%$ for the TMD-b parametrization, while for the TMD-a parametrization it is around 3$\%$ in all the figures of $J/\psi + \mathrm{jet}$ production.

\begin{figure}[H]
    \centering
    \begin{subfigure}{0.48\textwidth}
        \centering
        \includegraphics[width=\linewidth]{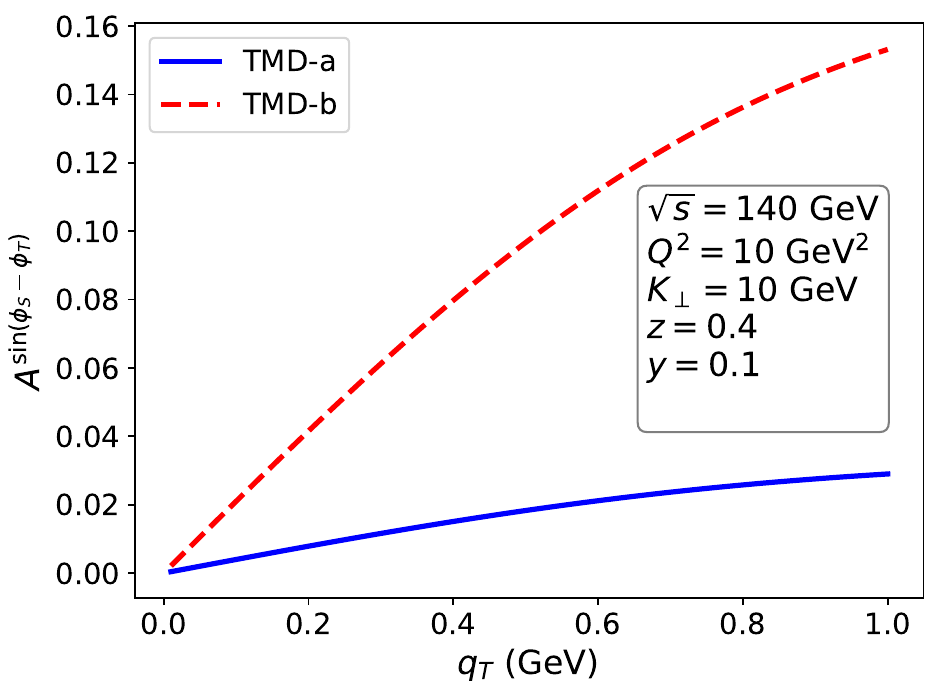}
        \caption{}
    \end{subfigure}
    \hfill
    \begin{subfigure}{0.48\textwidth}
        \centering
        \includegraphics[width=\linewidth]{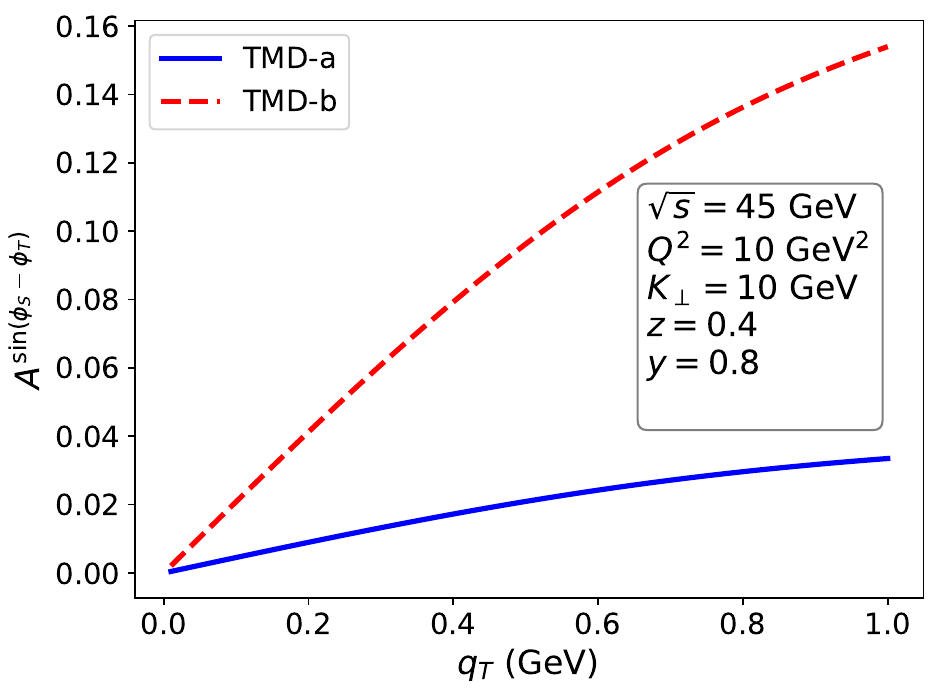}
        \caption{}
    \end{subfigure}
    \caption{\label{fig:4}
    Sivers asymmetry in $e+p^\uparrow\rightarrow e+J/\psi+\gamma +X$ process as a function of $q_T$  at the EIC kinematics $\sqrt{s}=140$ GeV (left) and $\sqrt{s}=45$   GeV (right). The asymmetry is estimated at fixed values of $Q^2=10~\text{GeV}^2$, $K_\perp=10~\text{GeV}$, $z=0.4$, and $y=0.1$ (for 140 GeV) and $y=0.8$ (for 45 GeV).}
\end{figure}

In Fig.~\ref{fig:4}, the Sivers asymmetry, $A_N^{\sin(\phi_S-\phi_T)}$, is shown as a function of $q_\sT$ for the $J/\psi+\gamma$ production.  The asymmetry is estimated at $\sqrt{s}=140$ GeV and 45 GeV, which is shown in Figs.~\ref{fig:4}(a) and (b), respectively. From Fig.~\ref{fig:4}, it is obvious that the asymmetry is almost independent of $\sqrt{s}$ for the chosen kinematical values.

In Fig.~\ref{fig:5}, the Sivers asymmetry as a function of $q_\sT$ is shown for $J/\psi+\mathrm{jet}$ production. Figs.~\ref{fig:5}(a) and (b) represent the asymmetry estimated at $\sqrt{s}=140$ GeV for BK11 and SV13 LDME sets, respectively, while  Figs.~\ref{fig:5}(c) and (d)  show the asymmetry at $\sqrt{s}=45$ GeV. In this case as well, the asymmetry shows very little dependence on $\sqrt{s}$.
The $q_\sT$-dependence of the asymmetry arises only from the Bessel functions. It can be expressed as 
\begin{align}
A_N(q_\sT) = -\frac{1}{2}\,\frac{\displaystyle N_g(x)\,\int \d b_\sT \, b_\sT^2\,J_1(b_\sT q_\sT)\, F_1(b_\sT)}{\displaystyle \int \d b_\sT \, b_\sT\,J_0(b_\sT q_\sT)\, F_2(b_\sT)} \,,\label{qt_ratio}
\end{align}
where $F_1(b_\sT)$ and $F_2(b_\sT)$ contain the gluon PDF, the Sudakov factor, and the hard part. 
Since the longitudinal momentum fraction $x$ is fixed by the kinematics of the process, $N_g(x)$ remains constant.\\
For small $q_\sT$, the Bessel functions can be approximated as
\begin{equation}
    J_1(b_\sT q_\sT) \simeq \frac{b_\sT q_\sT}{2}, \qquad
    J_0(b_\sT q_\sT) \simeq 1.
\end{equation}
Using these approximations, the asymmetry behaves as
\begin{align}
    A_N(q_\sT) \simeq -C\, N_g(x)\, q_\sT, \label{asym}
\end{align}
where $C$ is a positive constant determined by the ratio of the $b_\sT$ integrals in Eq.~\ref{qt_ratio}, including the factor 1/2.

Since $N_g(x)$ is negative in the present kinematical region, the overall negative sign in Eq.~\eqref{asym} leads to a positive asymmetry. As a result, in the small-$q_\sT$ region, the asymmetry increases approximately linearly with $q_\sT$.

\begin{figure}[H]
    \centering
    \begin{subfigure}{0.48\textwidth}
        \centering
        \includegraphics[width=\linewidth]{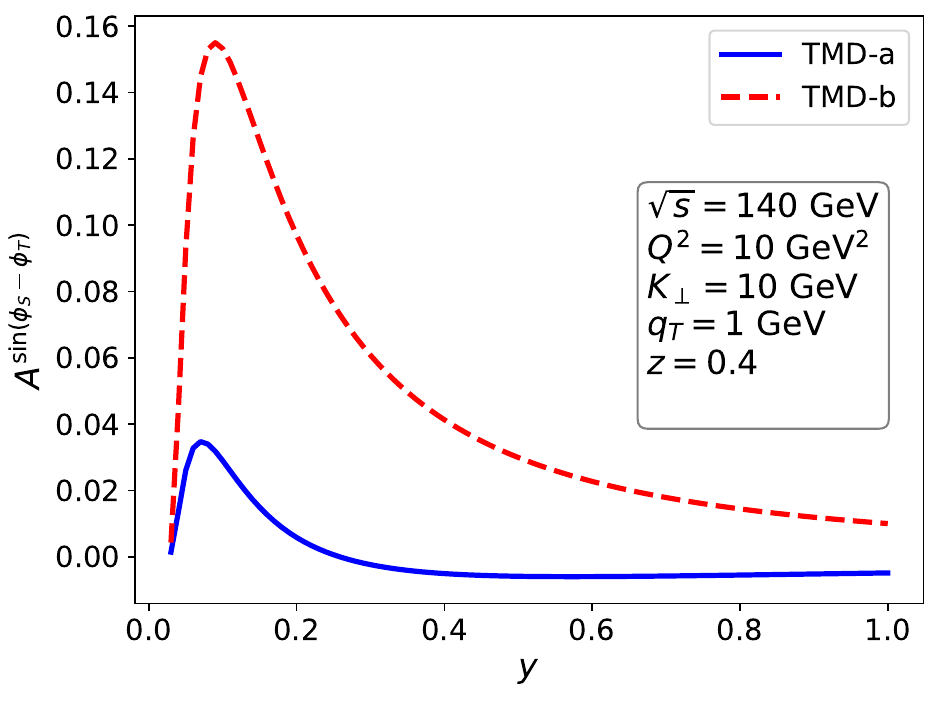}
        \caption{}
    \end{subfigure}
    \hfill
    \begin{subfigure}{0.48\textwidth}
        \centering
        \includegraphics[width=\linewidth]{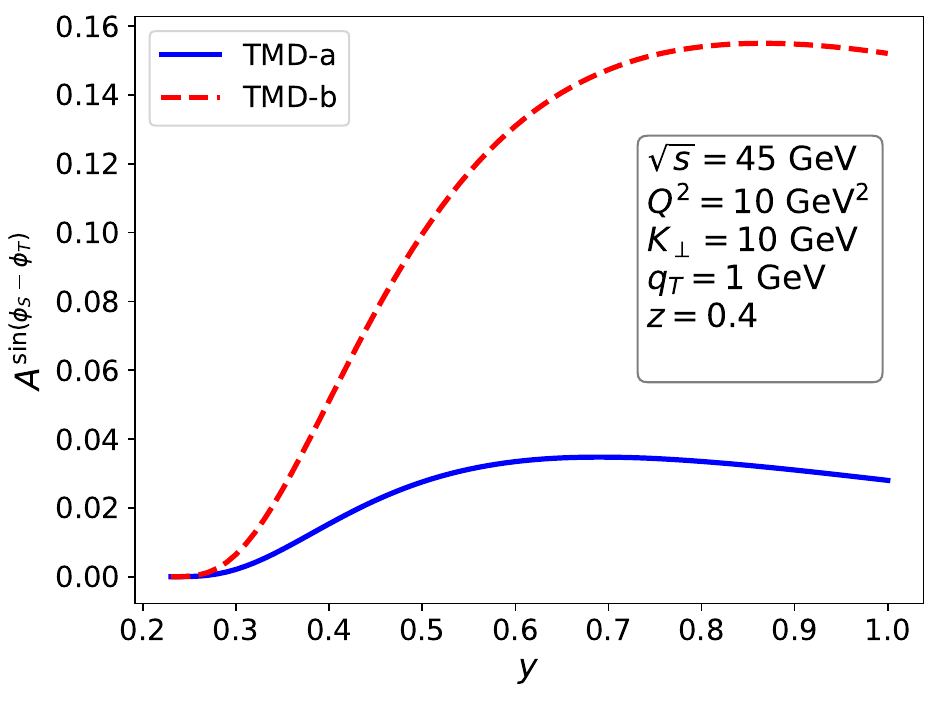}
        \caption{}
    \end{subfigure}
    \caption{\label{fig:6}
 Sivers asymmetry in $e+p^\uparrow\rightarrow e+J/\psi+\gamma +X$ process as a function of $y$  at the EIC kinematics $\sqrt{s}=140$ GeV (left) and $\sqrt{s}=45$ GeV (right). The asymmetry is estimated at fixed values of $Q^2=10~\text{GeV}^2$, $K_\perp=10~\text{GeV}$,  $q_\sT=1$ GeV, and $z=0.4$.}
\end{figure}

In our calculation, the Sivers asymmetry is found to be positive, in contrast to a few earlier studies in similar processes where it was found to be  negative. In particular, previous studies of $J/\psi + \gamma$ \cite{Chakrabarti:2022rjr} in electroproduction using Gaussian parametrization of gluon Sivers function and without TMD evolution, as well as $J/\psi + \text{jet}$ \cite{Kishore:2019fzb} in photoproduction with TMD evolution, both reported a negative Sivers asymmetry. This difference is due to different definitions of the Sivers function. In our work, the asymmetry is written directly in terms of $f_{1T}^{\perp g}(x, q_\sT)$. However, in previous studies, the asymmetry is expressed in terms of $\Delta^N f_{g/p^\uparrow}(x, q_\sT)$, which is related to $f_{1T}^{\perp g}(x, q_\sT)$ as per the Trento convention \cite{Bacchetta:2004jz}
\begin{align*}
\Delta^N f_{g/p^\uparrow}(x, q_\sT)= - \frac{2 q_\sT }{M_p}\,f_{1T}^{\perp g}(x, q_\sT).
\end{align*}
This minus sign in the definition of Sivers function leads to opposite results compared to our earlier work~\cite{Chakrabarti:2022rjr,Kishore:2019fzb}. \\
In Figs.~\ref{fig:6} and \ref{fig:7}, the asymmetry $A_N^{\sin(\phi_S-\phi_T)}$ is shown as a function of $y$ for $J/\psi+\gamma$ and $J/\psi+\mathrm{jet}$ production, respectively, at $Q^2=10~\text{GeV}^2$, $K_\perp=10~\text{GeV}$, $q_\sT=1~\mathrm{GeV}$, and $z=0.4$, and for two cm energies 140 \text{GeV} and 45 \text{GeV}. For $\sqrt{s}= 140 ~\text{GeV}$ (Figs.~\ref{fig:6}(a) and \ref{fig:7}(a)), as $y$ increases, the asymmetry first rises sharply, attains a maximum, and then decreases rapidly with increasing $y$. In contrast, for $\sqrt{s}= 45~\text{GeV}$ (Figs.~\ref{fig:6}(b) and \ref{fig:7}(b)), the asymmetry increases more gradually and neither a pronounced maximum nor a significant decrease is observed within the plotted range. 

The $y$-dependence of the Sivers asymmetry is qualitatively understood by $x$-dependence of the Sivers function, where the momentum fraction $x$ depends inversely on both $y$ and cm energy as given in Eq.~\eqref{del}.
\begin{align}
    x=\frac{K}{y s}, \qquad K=\text{constant.}
\end{align}
From Eq.~\eqref{Ng}, the $x$-dependence of the Sivers function is $N_g(x) \sim x^\alpha (1-x)^\beta$. In the low-$y$ region, the $x$ is large; hence, the $(1-x)^\beta$ factor becomes small, which suppresses the asymmetry. While, in the large-$y$ region, $x$ becomes small, making $x^\alpha$ small and again reducing the asymmetry. Therefore, the asymmetry rises at intermediate $y$, reaches a peak, and then starts decreasing. The cm energy $s$ strongly affects the position of the peak. For larger $s$, a fixed $x$ corresponds to a smaller $y$, so the peak appears at smaller $y$ and is clearly visible. For smaller $s$, the same $x$ corresponds to larger $y$, and the peak may not be visible within the kinematic range.

\begin{figure}[H]
    \centering
    \begin{subfigure}{0.48\textwidth}
        \centering
        \includegraphics[width=\linewidth]{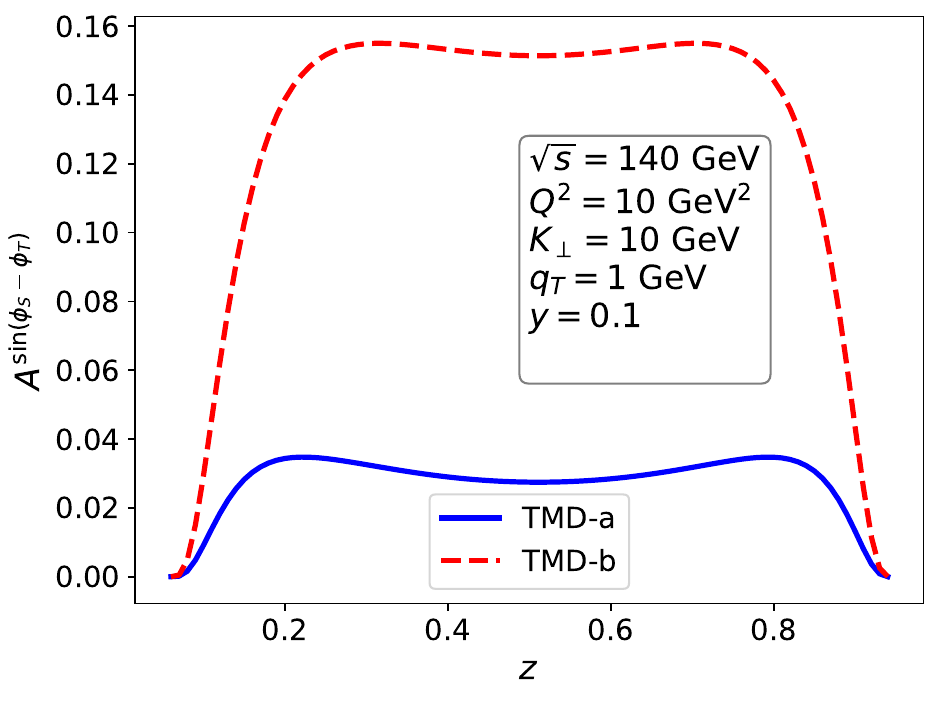}
        \caption{}
    \end{subfigure}
    \hfill
    \begin{subfigure}{0.48\textwidth}
        \centering
        \includegraphics[width=\linewidth]{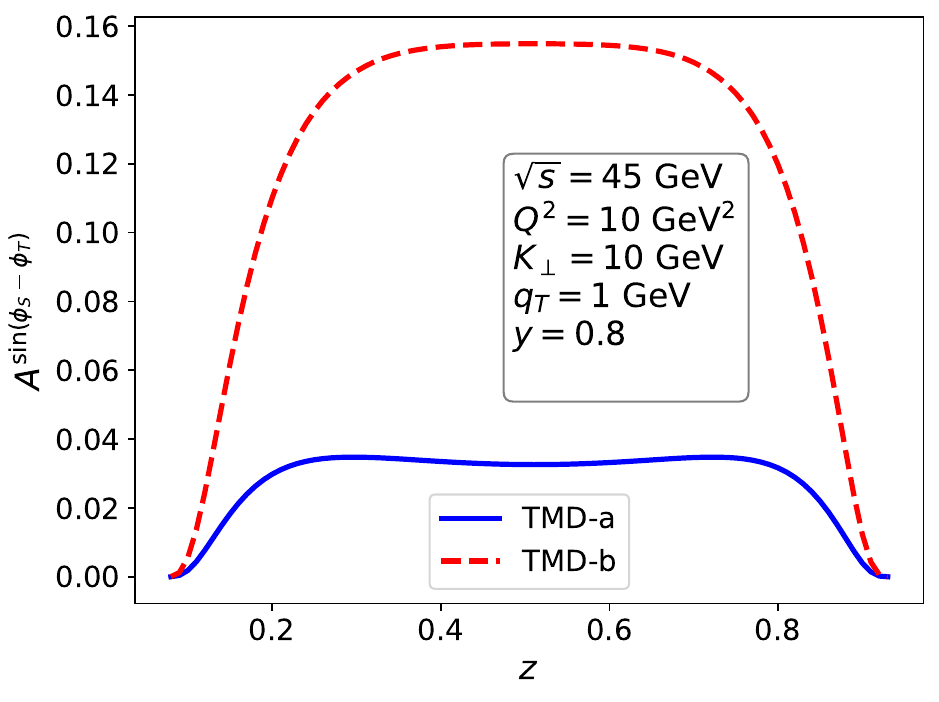}
        \caption{}
    \end{subfigure}
    \caption{\label{fig:8}
Sivers asymmetry in $e+p^\uparrow\rightarrow e+J/\psi+\gamma +X$ process as a function of $z$  at the EIC kinematics $\sqrt{s}=140$ GeV (left) and  $\sqrt{s}=45$ GeV (right). The asymmetry is estimated at fixed values of $Q^2=10~\text{GeV}^2$, $K_\perp=10~\text{GeV}$, $q_\sT=1$ GeV, and $y=0.1$ (for 140 GeV) and $y=0.8$ (for 45 GeV).}
\end{figure}

In Figs.~\ref{fig:8} and \ref{fig:9}, we present $A_N^{\sin(\phi_S-\phi_T)}$ as a function of $z$ at $Q^2=10~\text{GeV}^2$, $K_\perp=10~\text{GeV}$, $q_\sT=1~\mathrm{GeV}$, and $y=0.1$ (for 140 GeV) and $y=0.8$ (for 45 GeV) for $J/\psi+\gamma$ and $J/\psi+\mathrm{jet}$ production, respectively. Figs.~\ref{fig:8}(a) and~\ref{fig:9}(a) correspond to $\sqrt{s}=140~\text{GeV}$, while Figs.~\ref{fig:8}(b) and~\ref{fig:9}(b) correspond to $\sqrt{s}=45~\text{GeV}$.
At smaller values of $z$, the asymmetry is very small for both processes. As $z$ increases, the asymmetry increases in both cases. In the intermediate-$z$ region, the asymmetry for the $J/\psi+\gamma$ process remains nearly constant with $z$, whereas for the $J/\psi+\mathrm{jet}$ process, it decreases approximately linearly with $z$. For larger values of $z$, the asymmetry decreases again for both processes. This qualitative behaviour is observed for both cm energies.
The $z$-dependence of the Sivers asymmetry is governed by the $x$-dependence of the Sivers function and the unpolarized TMD, together with the hard part. The relation between the momentum fraction $x$ and $z$ is given in  Eq.~\eqref{del}
\begin{align}
x(z) = \frac{(1-z)(M_\psi^2 + \bm P_{\psi\perp}^2) + z \bm p_{a'\perp}^2 + z(1-z) Q^2}{z(1-z)ys}\,.
\end{align}
From Eq.~\eqref{Ng}, the $x$-dependence of the Sivers function is $N_g(x) \sim x^\alpha (1-x)^\beta$. At very small and large values of $z$, the momentum fraction $x$ is large, so $(1-x)^\beta$ becomes small, leading to a small asymmetry. As $z$ increases from small values, $x$ decreases. In this region, the enhancement due to $(1-x)^\beta$ dominates over the suppression due to $x^\alpha$, leading to increase in the asymmetry. However, in the intermediate-$z$ region, the two processes show different behaviours. The Sivers function and the unpolarized TMD are nearly constant in this region because the longitudinal momentum fraction $x$ varies very slowly with $z$. Therefore, for the $J/\psi+\mathrm{jet}$ process, the observed $z$-dependence of the asymmetry is mainly driven by the hard-scattering amplitudes, leading to an approximately linear decrease with $z$. On the other hand, for the $J/\psi+\gamma$ process, the asymmetry becomes almost flat in this region, since the hard part dependence cancels between the numerator and denominator, resulting in only a weak dependence on $z$. As $z$ increases further, $x$ starts to increase again, and the suppression of $(1-x)^\beta$ becomes dominant, causing $N_g(x)$ and the asymmetry to decrease.
\section{Conclusion}
\label{sec:conclusion}
In this work, we have estimated the Sivers asymmetry in $J/\psi$-photon and $J/\psi$-jet electroproduction, in the EIC kinematics, considering almost back-to-back production of the outgoing particles in the transverse plane. This kinematics allows us to use the TMD factorization. The NRQCD framework is used for the quarkonium formation.  In  $J/\psi$-photon production process, the contribution comes only from the CO $^3S_1^{(8)}$ state, whereas the $J/\psi$-jet production includes both CS and CO contributions. We have obtained a significantly larger cross section in $J/\psi$-jet process than in $J/\psi$-photon process. In this work, we have mainly studied the role of TMD evolution on the cross section and asymmetry. In the $J/\psi$-photon process, we observed that the asymmetry is independent of the LDME due to its exact cancellation between the numerator and denominator. On the other hand, in the $J/\psi$-jet process, the asymmetry shows only a weak LDME dependence due to the dominance of the gluon channel over the quark channel. Our result shows a sizable asymmetry for both the considered processes, which can serve as a promising experimental probe at the future EIC for accessing the gluon Sivers function.
\newpage
\bibliographystyle{apsrev}
\bibliography{reference}
\end{document}